\documentclass[lettersize,journal]{IEEEtran}

\usepackage{cite}
\usepackage{amsmath}
\usepackage{amsmath,amssymb,amsfonts}
\usepackage{algorithmic}
\usepackage{graphicx}
\usepackage{textcomp}
\usepackage{xcolor}
\usepackage{pdfpages}
\usepackage{textcomp}
\usepackage{enumerate}
\usepackage{xfrac}
\usepackage{amssymb}
\usepackage{mathtools}
\usepackage{amsmath}
\usepackage{enumerate}
\usepackage{blindtext}
\usepackage{enumitem}
\usepackage{graphicx}
\usepackage{caption}
\usepackage{subcaption}
\graphicspath{{Figures/}}
\usepackage[justification=centering]{caption}
\usepackage{setspace}
\usepackage{bbm}
\usepackage{bm}
\usepackage{optidef}
\usepackage {comment} 
\usepackage{listings}
\usepackage{pgfplots}
\usepackage{pgfplotstable}
\pgfplotsset{compat=newest}
\pgfplotsset{compat=1.15}
\usepackage{xcolor}
\pgfplotsset{every tick label/.append style={font=\tiny}}

\usepackage{hyperref}
\hypersetup{
	colorlinks=true,
	linkcolor=blue,
	filecolor=blue,   
	citecolor=blue,  
	urlcolor=blue,
	pdftitle={Deep Learning-Enabled Multi-Tag Detection in Ambient Backscatter Communications},
	pdfpagemode=FullScreen,
}
\usepackage{tikz}
\usepackage{amsmath}
\usepackage{amsfonts}
\usepackage{dblfloatfix}  
\usetikzlibrary{positioning}
\usetikzlibrary{positioning, shapes.geometric}
\usetikzlibrary{positioning, arrows.meta}
\usetikzlibrary{decorations.pathreplacing}

\DeclareMathOperator*{\argmin}{arg\,min}
\DeclareMathOperator*{\argmax}{arg\,max}

\definecolor{figcolor1}{RGB}{228,26,28} 
\definecolor{figcolor2}{RGB}{55,126,184}   
\definecolor{figcolor3}{RGB}{77,175,74}   
\definecolor{figcolor4}{RGB}{152,78,163}  
\definecolor{figcolor5}{RGB}{255,127,0}

\begin{document}
	
	\title{Deep Learning-Enabled Multi-Tag Detection in Ambient Backscatter Communications}
	\author{\IEEEauthorblockN{Talha Aky{\i}ld{\i}z and Hessam Mahdavifar\\} 	
		\thanks{T. Aky{\i}ld{\i}z is with the EECS Dept., University of Michigan, Ann Arbor, MI, 48104, USA (email: akyildiz@umich.edu).}
		\thanks{H. Mahdavifar is with the EECS Dept., University of Michigan, Ann Arbor, MI, 48104, USA and ECE Dept., Northeastern University, Boston, MA, 02115, USA (email: hessam@umich.edu).}
		\thanks{Part of this work was presented at the Asilomar Conference on Signals, Systems, and Computers 2025 \cite{akyildiz2025ambient}.} 
		
	}
	
	\maketitle
	\normalsize
	\begin{abstract}	
	Ambient backscatter communication (AmBC) enables battery-free connectivity by letting passive tags modulate existing RF signals, but reliable detection of multiple tags is challenging due to strong direct link interference, very weak backscatter signals, and an exponentially large joint state space. Classical multi-hypothesis likelihood ratio tests (LRTs) are optimal for this task when perfect channel state information (CSI) is available, yet in AmBC such CSI is difficult to obtain and track because the RF source is uncooperative and the tags are low-power passive devices. We first derive analytical performance bounds for an LRT receiver with perfect CSI to serve as a benchmark. We then propose two complementary deep learning frameworks that relax the CSI requirement while remaining modulation-agnostic. EmbedNet is an end-to-end prototypical network that maps covariance features of the received signal directly to multi-tag states. ChanEstNet is a hybrid scheme in which a convolutional neural network estimates effective channel coefficients from pilot symbols and passes them to a conventional LRT for interpretable multi-hypothesis detection. Simulations over diverse ambient sources and system configurations show that the proposed methods substantially reduce bit error rate, closely track the LRT benchmark, and significantly outperform energy detection baselines, especially as the number of tags increases.
	\end{abstract}

	\vspace{-0.4cm}
	\section{Introduction}
	Ambient backscatter communication is a pioneering technology that enables battery-free devices to communicate by leveraging existing radio frequency (RF) signals in the environment \cite{Huynh2017Ambient,Lu2018Ambient,niu2019overview}. This technique has gained significant attention in recent years due to its potential to enable ultra-low-power and ubiquitous communication for Internet of Things (IoT) devices \cite{atzori2010internet,rose2015internet,li2015internet}. The ability to operate without external power sources also enhances sustainability, making ambient backscatter a crucial component in advancing wireless communication technologies \cite{duan2020ambient,toro2021backscatter,akyildiz20206g}.
	
	Typically, an ambient backscatter system comprises three main components: an ambient RF source, a backscatter transmitter (tag), and a reader. The backscatter transmitter is a simple device equipped with an antenna and a switch that modulates the ambient RF signals received from the source by reflecting them in a way that encodes data, altering properties like amplitude, phase, or frequency \cite{chowdhury20206g}. The reader, often a more complex device, detects these modulations and decodes the transmitted data, by differentiating between the modulated backscatter signal and the original ambient signal \cite{liu2013ambient}. 
	
	Detecting signals from tags in ambient backscatter communication systems presents several difficulties due to the inherent properties of the technology. The primary difficulty lies in the weak nature of backscatter signals, which are significantly weaker than the ambient RF signals. The presence of noise and interference from other RF sources also exacerbates the challenge. The lack of precise channel state information at the reader also prevents effective signal decoding \cite{jiang2023backscatter}. However, accurate signal detection is still critical to ensure reliable data transmission between tags and readers and to allow AmBC systems to coexist with other wireless technologies in RF environments \cite{xu2023state}. New techniques such as advanced signal processing algorithms, machine and deep learning based detection under different configurations have been explored to address these challenges \cite{xu2018practical}. Specific techniques and approaches explored in the literature, mostly for the single tag detection problem, include differential encoding \cite{Wang2016Ambient}, semi-coherent detection \cite{Qian2016Semi-Coherent}, maximum likelihood non-coherent detection \cite{Qian2017Noncoherent}, cooperative communication \cite{Yang2018Cooperative}, non-coherent detection with OFDM signals \cite{Darsena2019Noncoherent}, ternary coded signaling \cite{Liu2017Coding}, and backscatter modulation with OFDM \cite{Elmossallamy2019Noncoherent}, among others. However, much less attention is given to the detection problem for multiple tags. In one prior work, \cite{Yang2023Non-Coherent} proposed an expectation maximization approach to cluster received signals and decode for the multi-tag detection problem. 
	
	\begin{table*}[t]
	\caption{Comparison with representative AmBC detection methods}
	\vspace{-0.1cm}
	\label{tab:comparison}
	\centering
	\setlength{\tabcolsep}{3pt}
	\renewcommand{\arraystretch}{1.35}
	\scriptsize
	
	\begin{tabular}{p{1.95cm}|p{1.65cm}|p{1.8cm}|p{2.05cm}|p{2.4cm}|p{1.99cm}|p{4.8cm}}
	\hline
	\textbf{Work} & \textbf{Tags} & \textbf{Hypothesis space} & \textbf{CSI requirement} & \textbf{Runtime adaptation} & \textbf{Modulation} & \textbf{Key characteristics} \\ \hline

	Liu et al. \cite{liu2021deep} & Single ($N=1$) & 2 hypotheses & CSI-free (covariance features) & Required (online fine-tuning) & Gaussian and modulated ambient sources & Deep transfer learning based LRT with a covariance matrix neural network. Single tag binary detector with pilot-based online fine-tuning. \\ \hline

	Liu et al. \cite{liu2020deep} & Single ($N=1$) & 2 hypotheses & CSI from pilots via deep residual CNN & None (fixed weights after training) & Gaussian ambient source only & Deep residual CNN for pilot-based channel estimation in a single tag system. Used as a CSI acquisition module. \\ \hline

	Yang et al. \cite{Yang2023Non-Coherent} & Multi ($N\ge2$) & $2^N$ hypotheses & No CSI (non-coherent detection) & None (training-free) & FM0/Miller coded tags with modulated ambient sources & Expectation maximization clustering with state flipping waveform structure, asynchronous multi-tag detection without training. \\ \hline

	\textbf{EmbedNet} (ours) & Multi ($N\ge2$) & $2^N$ hypotheses & CSI-free (prototype based) & None (fixed neural network with prototype update only) & Gaussian and modulated ambient sources & Prototypical $2^N$-way detector where pilots define prototypes in a learned embedding space. Multi-tag detection without weight updates.  \\ \hline

	\textbf{ChanEstNet} (ours) & Multi ($N\ge2$) & $2^N$ hypotheses & Implicit CSI from pilots & None (fixed weights after training, pilots only affect LRT inputs) & Gaussian and modulated ambient sources & Channel estimates from CNN followed by multi-hypothesis LRT. Provides explicit CSI and near-LRT detection without online fine-tuning. \\ \hline
	\end{tabular}
	\vspace{-0.6cm}
	\end{table*}

	Machine learning (ML) and deep learning (DL) techniques have been applied to enhance single tag signal detection in AmBC systems, offering advantages over traditional methods through data-driven adaptability to complex signal environments \cite{Zhang2019Constellation, Hu2019Machine,Zhang2017Signal,Wang2019Machine,Guo2019Exploiting,liu2020deep,liu2021deep}. Specifically, \cite{Hu2019Machine} proposed transforming signal detection into a classification task to use supervised algorithms, while an unsupervised learning based detection method is proposed in \cite{Zhang2017Signal}. Other major approaches include statistical clustering framework \cite{Guo2019Exploiting}, deep residual learning approach \cite{liu2020deep}, deep transfer learning (DTL) \cite{liu2021deep}, conditional generative adversarial networks (CGANs) \cite{zargari2024deep}, and deep neural network with CSI acquisition \cite{zargari2024enhancing}. Other studies \cite{jameel2020low, Jia2021IRS-Assisted, akyildiz2022ml} explored deep learning for signal detection in specific applications involving AmBC, including intelligent reflecting surfaces (IRS)-assisted AmBC, reconfigurable intelligent surfaces (RIS)-assisted AmBC, wireless-powered AmBC, and UHF RFID systems. A comprehensive review of detection techniques and applications for AmBC networks is provided in \cite{toro2021backscatter, zargari2023signal}. Again, these prior works mostly focus on the single tag detection problem. DL-enabled multi-tag detection has been explored in the context of UHF RFID systems in \cite{akyildiz2022ml}, which can be considered as a special case of AmBC with one device serving as both the RF source and the reader.

	In this paper, we focus on multi-tag detection in ambient backscatter communication (AmBC) systems and develop two complementary deep learning based receivers that explicitly address the exponential $2^N$ hypothesis space\footnote{$N$ denotes the number of tags. The joint tag state space consists of $2^N$ distinct combinations since each tag has two possible reflection states.}. Existing deep learning methods for AmBC are either limited to single tag detection or rely on assumptions that do not scale to multiple tags. The deep transfer learning framework in \cite{liu2021deep} targets a single tag (binary decisions) and requires online fine-tuning of network weights, which incurs inference time complexity and does not extend to joint multi-tag states. The approach in \cite{liu2020deep} uses deep residual learning for channel estimation but is also restricted to single tag systems and does not directly perform multi-tag detection. For multi-tag scenarios, the non-coherent scheme in \cite{Yang2023Non-Coherent} employs expectation maximization (EM) together with FM0 line coding, but it is tied to specific coding structures and lacks the modulation-agnostic flexibility.

	In contrast, we introduce two deep learning frameworks that relax the need for explicit CSI while remaining modulation-agnostic and scalable with the number of tags. EmbedNet is an end-to-end prototypical network that performs $2^N$-way classification from covariance features, adapting to each frame via pilot-based prototypes without any weight updates during inference. ChanEstNet is a hybrid design in which a specialized convolutional neural network (CNN) refines pilot correlations into effective channel coefficients, which are then fed to a classical multi-hypothesis LRT for interpretable detection. Both approaches avoid transfer learning or online fine-tuning, operate across different ambient source modulations, and are designed for multi-tag operation. Table \ref{tab:comparison} summarizes the main differences in modeling assumptions, hypothesis space, and key characteristics across these methods. The contributions of this work are summarized as follows:

	{
	\begin{enumerate}
		\item We formulate the multi-tag AmBC detection problem with an exponentially large joint state space and derive analytical benchmarks, including a multi-hypothesis LRT with perfect CSI, pairwise error probability bounds, and an energy detector, to characterize the fundamental performance limits.

		\item We propose EmbedNet, an end-to-end prototypical network that operates on covariance features and performs $2^N$-way multi-tag detection via pilot-based prototypes, enabling per frame adaptation without CSI knowledge.

		\item We develop ChanEstNet, a hybrid receiver in which a specialized channel estimation network maps pilot correlations to effective channel coefficients that are then used by a classical multi-hypothesis LRT, combining data-driven channel estimation with classical theory.

		\item We carry out extensive simulations under diverse conditions, showing that both EmbedNet and ChanEstNet closely track the LRT benchmark, and remain robust across different system configurations.
	\end{enumerate}
	}

	The remainder of this paper is organized as follows. Section II presents the system model, providing a detailed description of the AmBC system with multiple tags. Section III explores the theoretical limits of AmBC with likelihood ratio test, pairwise error probability bound, and energy detector, establishing the performance bounds. Section IV introduces our deep learning based signal detection approaches. Section V presents comprehensive numerical results from our simulations. Finally, Section VI concludes the paper.

	Throughout this work, the notation \(\mathbb{E}[\cdot]\) denotes the expectation of a random variable, while \(\operatorname{Re}(\cdot)\) and \(\operatorname{Im}(\cdot)\) extract the real and imaginary parts of a complex valued quantity. The symbol \(\lvert x\rvert\) refers to the magnitude of a complex number. For a vector \(\mathbf{x}\), the notation \(\|\mathbf{x}\|\) indicates its Euclidean norm, and \(\|\mathbf{x}\|^2 = \mathbf{x}^\mathsf{H}\mathbf{x}\) is its squared norm where \(\mathsf{H}\) denotes the conjugate transpose operation. For a complex scalar \(x\), \(x^*\) denotes complex conjugation. For matrices, \(\det(\mathbf{A})\) is the determinant of \(\mathbf{A}\), and \(\mathbf{A}^{-1}\) denotes the inverse of \(\mathbf{A}\).	
	\begin{figure}[t]
		\centering
		\resizebox{1\linewidth}{!}{
		\begin{tikzpicture}[>=stealth,scale=1.0]
			\node[draw, fill=figcolor3!60, text centered, text width=1.6cm, minimum height=0.8cm, font=\normalsize] (RFSource) at (0, 3.1) {RF Source};
			
			\draw[thick] (-0.3,3.5) -- (0.3,3.5);
			\draw[thick] (-0.2,3.7) -- (0.2,3.7);
			\draw[thick] (-0.1,3.9) -- (0.1,3.9);
			\draw[thick] (0,3.5) -- (0,4);
			\node[text width=2.8cm, font=\scriptsize, align=center] at (0, 4.3) {Ambient Signal $s_k^{(t)}$};
			
			\node[draw, fill=figcolor2!60, text centered, text width=1.2cm, minimum height=0.6cm] (Tag1) at (5.5, 5) {Tag 1};
			\node[text width=2cm, font=\scriptsize] at (5.7, 5.5) {$c_1^{(t)} \in \{0,1\}$};
			
			\node[draw, fill=figcolor2!60, text centered, text width=1.2cm, minimum height=0.6cm] (Tag2) at (5.5, 3.5) {Tag 2};
			\node[text width=2cm, font=\scriptsize] at (5.7, 4) {$c_2^{(t)} \in \{0,1\}$};
			
			\node[draw, fill=figcolor2!60, text centered, text width=1.2cm, minimum height=0.6cm] (TagN) at (5.5, 1.5) {Tag N};
			\node[text width=2cm, font=\scriptsize] at (5.7, 2) {$c_N^{(t)} \in \{0,1\}$};
			
			\node at (5.5, 2.75) {$\vdots$};
			
			\node[draw, fill=figcolor1!60, text centered, text width=2.5cm, minimum height=0.8cm, font=\normalsize] (Reader) at (3, 0) {Reader};
			
			\draw[thick] (2.0,0.4) -- (2.0,0.9);
			\draw[thick] (2.3,0.4) -- (2.3,0.9);
			\draw[thick] (2.6,0.4) -- (2.6,0.9);
			\node at (3.0, 0.7) {$\ldots$};
			\draw[thick] (3.4,0.4) -- (3.4,0.9);
			\draw[thick] (3.7,0.4) -- (3.7,0.9);
			\draw[thick] (4.0,0.4) -- (4.0,0.9);
			
			\node[text width=3cm, text centered, font=\scriptsize] at (3,-0.6) {$M$-element antenna array};
			
			\draw[->, figcolor2!80, thick] (RFSource) -- node[near start, above, sloped, font=\scriptsize] {$f_1$} (Tag1.west);
			\draw[->, figcolor2!80, thick] (RFSource) -- node[near start, above, sloped, font=\scriptsize] {$f_2$} (Tag2.west);
			\draw[->, figcolor2!80, thick] (RFSource) -- node[near start, above, sloped, font=\scriptsize] {$f_N$} (TagN.west);
			
			\draw[->, figcolor1!80, thick] (Tag1.west) -- node[very near start, above, sloped, font=\scriptsize] {$\mathbf{g_1}$} (Reader.north);
			\draw[->, figcolor1!80, thick] (Tag2.west) -- node[very near start, above, sloped, font=\scriptsize] {$\mathbf{g_2}$} (Reader.north);
			\draw[->, figcolor1!80, thick] (TagN.west) -- node[very near start, above, sloped, font=\scriptsize] {$\mathbf{g_N}$} (Reader.north);
			
			\draw[->, figcolor3!80, thick] (RFSource) -- node[midway, left, font=\scriptsize] {$\mathbf{h}$} (Reader.north);
			
			\draw[figcolor2!80, thick] (-0.5,-1) -- (0,-1);
			\node[right, font=\scriptsize] at (0,-1) {Forward channel};
			
			\draw[figcolor1!80, thick] (2,-1) -- (2.5,-1);
			\node[right, font=\scriptsize] at (2.5,-1) {Backscatter channel};
			
			\draw[figcolor3!80, thick] (4.7,-1) -- (5.2,-1);
			\node[right, font=\scriptsize] at (5.2,-1) {Direct channel};
		\end{tikzpicture}}
		\vspace{-0.5cm}
		\caption{Multi-tag AmBC system with a single antenna RF source, $N$ backscatter tags, and an $M$-antenna reader.}
		\label{fig1}
		\vspace{-0.5cm}
	\end{figure}
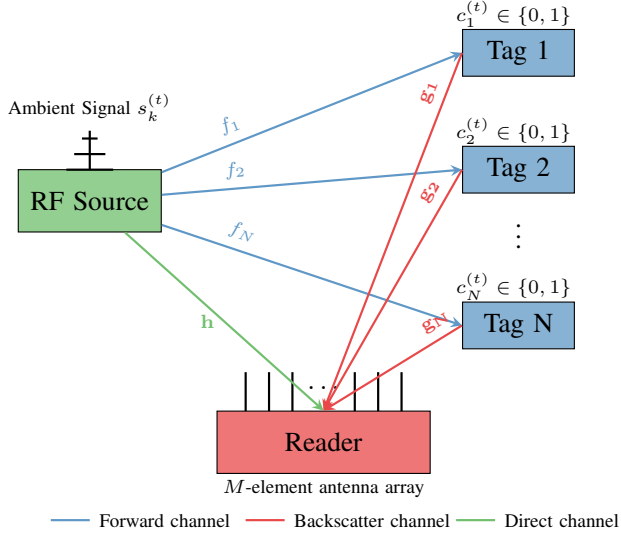

	\section{System Model}
	The multi-tag AmBC system consists of an ambient RF source, $N$ passive tags ($N \geq 1$), and a single reader equipped with an $M$-element antenna array for signal detection where ambient RF source and passive tags equipped with single antenna as shown in Fig. \ref{fig1}. The ambient RF source continuously broadcasts a waveform, which is not under direct control of the reader or the tags. In practice, it may be a cellular base station, a Wi-Fi access point, or another uncooperative transmitter. Each of the $N$ tags encodes data by modulating its load impedance. If the $i$-th tag is in the reflective state, it backscatters the ambient signal, otherwise it effectively absorbs the incident signal. We follow the on/off keying (OOK) modulation approach commonly used in AmBC systems. The reader, in turn, observes the received signal at each of the $M$ antennas which is a superposition of the direct path signal from the RF source plus the backscattered signals from all $N$ tags. Its objective is to detect tag bits under strong direct path and limited or unknown channel conditions.
	
\begin{figure}[t]
		\centering
		\resizebox{1\linewidth}{!}{
		\begin{tikzpicture}[scale=0.6]
			\tikzset{
				box1/.style={draw, fill=figcolor1!60, minimum width=1cm, minimum height=0.4cm, font=\footnotesize},
				box2/.style={draw, fill=figcolor2!60, minimum width=1cm, minimum height=0.4cm, font=\footnotesize},
				box3/.style={draw, fill=figcolor3!60, minimum width=1cm, minimum height=0.4cm, font=\footnotesize},
				tagbox1/.style={draw, fill=figcolor1!60, minimum width=1cm, minimum height=0.4cm, font=\small},
				tagbox2/.style={draw, fill=figcolor2!60, minimum width=1cm, minimum height=0.4cm, font=\small},
				tagbox3/.style={draw, fill=figcolor3!60, minimum width=1cm, minimum height=0.4cm, font=\small},
				node distance=0.5cm
			}
			
			\node[tagbox1] (tag1) at (-3.5,2) {Tag 1};
			\node[tagbox2] (tag2) at (-3.5,0) {Tag 2};
			\node[tagbox3] (tagN) at (-3.5,-3) {Tag N};
			
			\foreach \x [count=\i] in {1,2} {
				\pgfmathsetmacro{\xpos}{\i*1.65-3}
				\node[box1] (c1\i) at (\xpos,2) {$c_1^{\x}$};
				\node[box2] (c2\i) at (\xpos,0) {$c_2^{\x}$};
				\node[box3] (cN\i) at (\xpos,-3) {$c_N^{\x}$};
			}
			
			\foreach \x [count=\i] in {P,{P+1}} {
				\pgfmathsetmacro{\xpos}{\i*1.65+1.5}
				\node[box1] (c1\i) at (\xpos,2) {$c_1^{\x}$};
				\node[box2] (c2\i) at (\xpos,0) {$c_2^{\x}$};
				\node[box3] (cN\i) at (\xpos,-3) {$c_N^{\x}$};
			}
			
			\foreach \x [count=\i] in {{T-1},T} {
				\pgfmathsetmacro{\xpos}{\i*1.65+6}
				\node[box1] (c1\i) at (\xpos,2) {$c_1^{\x}$};
				\node[box2] (c2\i) at (\xpos,0) {$c_2^{\x}$};
				\node[box3] (cN\i) at (\xpos,-3) {$c_N^{\x}$};
			}
			
			\foreach \y in {2,0,-3} {
				\node at (1.75,\y) {$\cdots$};
				\node at (6.25,\y) {$\cdots$};
			}
			
			\node at (-3.5, -1.25) {$\vdots$};
			
			\draw [decoration={brace,mirror,raise=5pt},decorate] (-2.25,-3.5) -- (3.75,-3.5) 
			node [pos=0.5,below=10pt] {\small Pilot Symbols};
			\draw [decoration={brace,mirror,raise=5pt},decorate] (4,-3.5 ) -- (10.25,-3.5) 
			node [pos=0.5,below=10pt] {\small Data Symbols};
		\end{tikzpicture}}
		\caption{Frame structure of $N$ tags in AmBC, comprising $P$ pilot and $(T - P)$ data symbols.}
		\label{fig2}
		\vspace{-0.7cm}
	\end{figure}
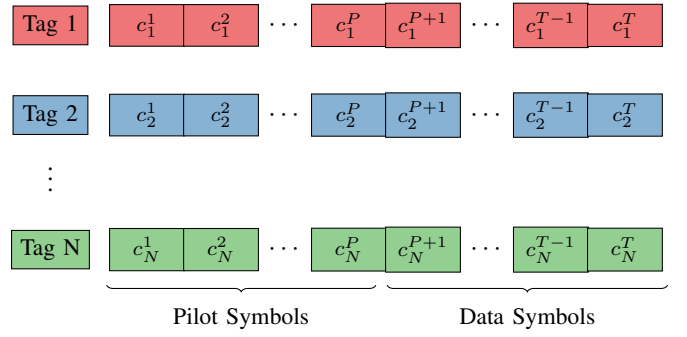

	We adopt a frame based transmission where each frame is composed of $T$ tag symbol periods, as depicted in Fig. \ref{fig2}. For a frame with $T$ periods, $P$ of them are designated as pilot symbols, while the remaining ($T-P$) periods carry data (information) bits for all tags. The pilot symbols, known a priori by the reader for all tags, enable initial synchronization, and facilitate channel inference. All tags are assumed to be synchronized frame-wise and to use the same frame length $T$.

	We first define $c_i^{(t)} \in \{0,1\}$ to denote the binary state of the $i$-th tag during the $t$-th symbol period, for $i = 1, 2,\dots,N$ and $t = 1, 2, \dots,T$. Each tag symbol $c_i(t)$ lasts for a fixed duration, and all tags are assumed to share the same symbol timing. Let $K$ be the number of ambient source symbols that occur within one tag symbol period. We refer to $K$ as the source-to-tag ratio (STR), since it reflects how many times faster the ambient source is transmitting relative to the bit rate of tags. We denote the ambient signal at sample index $k$ within the $t$-th tag period by $s_k^{(t)}$. The reflection state of $i$-th tag $c_i(t)$ remains constant throughout these $K$ source symbol intervals, we have $c_i^{(t)} = c_i^{(t+1)} = \dots = c_i^{(t+K-1)}, \ i = 1, 2, \dots, N$. 
	
	We now describe the channel model between RF source, $N$ passive tags and the reader with $M$ antenna arrays. Each tag $i$ receives the ambient waveform $s_k^{(t)}$ from the RF source through a forward channel which is denoted by $f_i \in \mathbb{C}$. After capturing this waveform, tag $i$ backscatters its own signal through a backscatter channel to the reader which is characterized by the vector $\mathbf{g}_i \in \mathbb{C}^{M \times 1}$. Independently, the reader also collects signals from the RF source via a direct channel and it is denoted by $\mathbf{h} \in \mathbb{C}^{M \times 1}$. The direct link is generally typically much stronger than the backscatter path. To reflect this, we define reflection coefficient for each tag $i$ and denote it by $\alpha_i$ where it is significantly smaller than $1$.
	
	After representing all necessary parameters and variables, we now derive the received signal vector at the reader for the $k$-th RF source sample within the $t$-th tag symbol period denoted by $\mathbf{x}_k^{(t)}$, i.e., 
	\begin{equation}
		\label{eq:rx_signal}
		\mathbf{x}_k^{(t)}=
		\underbrace{\mathbf{h}  s_k^{(t)}}_{\text{direct path}} +
		\underbrace{\sum_{i=1}^{N} \alpha_i  f_i  \mathbf{g}_i  s_k^{(t)}  c_i^{(t)}}_{\text{combined backscatter}} + \hspace{-0.2cm}
		\underbrace{\mathbf{u}_k^{(t)}}_{\text{AWGN noise}} \hspace{-0.3cm}, \quad \forall k, \quad \forall t,
	\end{equation}
	where we model the noise as additive white Gaussian noise (AWGN) and assume that $\mathbf{u}_k{(t)}$ is a circularly symmetric complex Gaussian random vector with zero mean and covariance $\sigma_u^2 \mathbf{I}_M$, i.e., $\mathbf{u}_k^{(t)} \sim \mathcal{CN}(\mathbf{0}, \sigma_u^2\mathbf{I}_M)$. In other words, each antenna experiences independent AWGN noise realizations and the noise is spatially uncorrelated across antennas. For each tag symbol period with $K$ RF source samples, an observation matrix $\mathbf{X}^{(t)} = [\mathbf{x}_1^{(t)}, \mathbf{x}_2^{(t)}, \dots, \mathbf{x}_{K}^{(t)}] \in \mathbb{C}^{M \times K}$ is defined to represent all samples within one tag symbol period.
	
	Given this model, we can define the average signal-to-noise ratio (SNR) for the direct path between the RF source and reader, and the backscatter path of $i$-th tag and the reader, denoted by $\text{SNR}$ and $\text{SNR}_{b,i}$, respectively, as 
	\begin{align}
		\text{SNR} &= \frac{\mathbb{E}[\|\mathbf{h}s_k^{(t)}\|^2]}{\mathbb{E}[\|\mathbf{u}_k^{(t)}\|^2]}, & {\text{SNR}}_{b,i} = \frac{\mathbb{E}[ \|\alpha_i f_i \mathbf{g}_i s_k^{(t)}\|^2]}{\mathbb{E}[\|\mathbf{u}_k^{(t)}\|^2]}.
	\end{align}

	The average relative coefficient between the direct and backscatter signal paths for the $i$-th tag can be defined as
	\begin{equation}
		{\zeta}_i = \frac{\mathbb{E}[\|\alpha_i f_i \mathbf{g}_i\|^2]}{\mathbb{E}[\|\mathbf{h}\|^2]},
	\end{equation}
	where ${\zeta}_i$ indicates the average relative strength of the backscatter path compared to the direct path for $i$-th tag.
	
	In our AmBC system model, we consider a flat fading channel model, where the channel characteristics remain constant over the duration of a frame. This assumption simplifies the analysis and is often reasonable for the environments with relatively slow channel variations. Additionally, frame-wise tag synchronization can be achieved by leveraging the periodic structure of ambient RF signals. Tags equipped with simple envelope detectors can identify known preambles or synchronization sequences in ambient sources and align their frame boundaries accordingly \cite{yang2017modulation,parks2014turbocharging}. Once synchronized, low-power crystal oscillators maintain coherence over the frame duration. Moreover, since the direct path is typically much stronger than the backscatter components, the reader can recover ambient timing and, when needed, estimate or decode the ambient symbols over a frame.
	
	\section{Theoretical Error Bounds of multi-tag AmBC system}
	In this section, we examine three analytical methods that characterize the performance of multi-tag AmBC. We first derive the exact maximum likelihood detector assuming perfect CSI using likelihood ratio test. While it provides a benchmark for minimal error probability, it entails exponential complexity in the number of tags. We also analyze pairwise error probability (PEP) bound by approximating the multi-hypothesis detection problem and evaluating pairwise errors, obtaining a union bound. Lastly, we develop an energy detection (ED) approach based on the received energy over a tag symbol. 
	
	\vspace{-0.5cm}
	\subsection{Optimal log-likelihood ratio test}
	\label{subsec:LRT}
	In this section, we investigate the theoretical bounds of multi-tag AmBC systems under different RF source signal distribution using optimal log-likelihood ratio test with perfect CSI. In our analysis, we extend the conventional single tag AmBC framework to a multi-tag system, where $N$ passive tags are present. Unlike prior works, e.g., \cite{liu2021deep}, where error bounds were derived specifically for a single tag, which naturally leads to a binary hypothesis test, our analysis is more comprehensive, considering a system with multiple $N$ tags. We have $2^N$ possible hypotheses, $H = \{H_0, H_1, ..., H_{(2^N-1)}\}$, where each hypothesis $H_j$ corresponds to a specific tag state combination, i.e., $(c_1^{(t)}, c_2^{(t)}, \dots, c_N^{(t)}) = (b_{j,1}, b_{j,2}, \ldots, b_{j,N})$, where $b_{j,i}$ is the $i$-th bit in the binary representation of hypothesis $j$. The received signal at time $t$ for the $k$-th RF source sample is
	\begin{equation} \label{eq1}
		\begin{split}
			\mathbf{x}_k^{(t)} & = \mathbf{h}s_k^{(t)} + \sum_{i=1}^N \alpha_i f_i \mathbf{g}_i s_k^{(t)}  b_{j,i} + \mathbf{u}_k^{(t)}, \\
			& = \mathbf{w}_j s_k^{(t)}  + \mathbf{u}_k^{(t)},
		\end{split}
	\end{equation}
	where $\mathbf{w}_j = \mathbf{h} + \sum_{i=1}^N \alpha_i f_i \mathbf{g}_i b_{j,i}$. The likelihood function under hypothesis $H_j$ will follow a multivariate complex Gaussian distribution, i.e., $\mathbf{x}_k^{(t)} \sim \mathcal{CN}(\mathbf{w}_j s_k^{(t)}, \sigma_u^2\mathbf{I}_M)$, can be written as
	\begin{align}\label{eq2}
		&p(\mathbf{x}_k^{(t)} | H_j) = \nonumber \\ & \frac{1}{(\pi\sigma_u^2)^M} \exp\left(-\frac{1}{\sigma_u^2}(\mathbf{x}_k^{(t)} - \mathbf{w}_j s_k^{(t)})^H (\mathbf{x}_k^{(t)} - \mathbf{w}_j s_k^{(t)})\right).
	\end{align}
	Moreover, since each tag symbol period consists of $K$ RF source samples, we can write the log-likelihood function denoted by $\mathcal{L}_j$ for hypothesis $H_j$ over the observation matrix,
	$\mathbf{X}^{(t)} = [\mathbf{x}_1^{(t)}, \mathbf{x}_2^{(t)}, \dots, \mathbf{x}_{K}^{(t)}]$, i.e., 
	\begin{equation}
		\mathcal{L}_j = \log p(\mathbf{X}^{(t)}| H_j) = \sum_{k=1}^K \log p(\mathbf{x}_k^{(t)} | H_j).
	\end{equation}
	
	In our AmBC system, the ambient source signal \(s_k^{(t)}\) can be modeled in different ways, leading to further simplifications of the likelihood ratio test. In the first scenario, \(s_k^{(t)}\) is assumed to be a complex Gaussian random variable with zero mean and variance \(\sigma_s^2\), i.e., \(s_k^{(t)} \sim \mathcal{CN}(0, \sigma_s^2)\). This assumption leads to a likelihood function based on the multivariate complex Gaussian distribution, i.e., the received signal vector can be modeled as  \(\mathbf{x}_k^{(t)} \sim \mathcal{CN}(\mathbf{0}, \mathbf{\Sigma}_j)\) where \(\mathbf{\Sigma}_j = \sigma_s^2 \mathbf{w}_j {\mathbf{w}_j}^H + \sigma_u^2\mathbf{I}_M\). Then, we can simplify eq. (\ref{eq2}) further as follows
	\begin{align}\label{eq3}
		p(\mathbf{x}_k^{(t)} | H_j) = \frac{1}{(\pi)^M \det(\mathbf{\Sigma}_j)} \exp\left(-(\mathbf{x}_k^{(t)})^H {\mathbf{\Sigma}_j}^{-1} \mathbf{x}_k^{(t)}\right).
	\end{align}
	and log-likelihood function \(\mathcal{L}_j\) becomes 
	\begin{align}
		\mathcal{L}_j &= \log p(\mathbf{X}^{(t)} | H_j) = -K \left( M\log(\pi) +\log(\det(\mathbf{\Sigma}_j)) \right) \nonumber \\ &-\sum_{k=1}^K \left((\mathbf{x}_k^{(t)})^H {\mathbf{\Sigma}_j}^{-1} \mathbf{x}_k^{(t)}\right).
	\end{align}
	
	Alternatively, \(s_k^{(t)}\) can be modeled as a modulated signal drawn from a known $M$-ary constellation set \(\mathcal{C} = \{C_1, C_2, ..., C_M\}\) under equal probability and a power of \(\sigma_s^2\), where each \(C_m \in \mathcal{C}\) represents a symbol in the modulation set, chosen with equal probability. In this case, \(s_k^{(t)}\) can only take on discrete values, and the received signal vector can be simplified and modeled as  \(\mathbf{x}_k^{(t)} \sim \mathcal{CN}(\mathbf{w}_j C_m, \sigma_u^2\mathbf{I}_M)\) under hypothesis $j$. Similarly, we can simplify eq. (\ref{eq2}), rewrite it as
	\begin{align}\label{eq4}
		&p(\mathbf{x}_k^{(t)} | H_j) = \nonumber \\ & \frac{1}{(\pi\sigma_u^2)^M} \exp\left(-\frac{1}{\sigma_u^2}(\mathbf{x}_k^{(t)} - \mathbf{w}_j C_m)^H (\mathbf{x}_k^{(t)} - \mathbf{w}_j C_m)\right).
	\end{align}
	and log-likelihood function \(\mathcal{L}_j\) becomes 
	\begin{align}
		\mathcal{L}_j &= \log p(\mathbf{X}^{(t)} | H_j) = -KM \log(\pi\sigma_u^2) \nonumber \\ & - \frac{1}{\sigma_u^2}\sum_{k=1}^K (\mathbf{x}_k^{(t)} - \mathbf{w}_j C_m)^H (\mathbf{x}_k^{(t)} - \mathbf{w}_j C_m).
	\end{align}
	
	Finally, in general terms, the log-likelihood ratio test and decision rule can be formulated as
	\begin{align}  
		{j}^* &= \argmax_{j \in \{0,1,\ldots,2^N-1\}} \log p(\mathbf{X}^{(t)}| H_j),
	\end{align}
	where \(H_{{j}^*}\) is the optimal hypothesis.
	
	\vspace{-0.25cm}
	\subsection{Pairwise Error Probability Analysis}
	The optimal likelihood ratio test normally tests all \(2^N\) hypotheses. To avoid this exponential complexity, we analyze the pairwise error probability (PEP) between a reference hypothesis \(H_0\) and an alternative hypothesis \(H_j\). Let the Hamming distance between the binary vectors corresponding to \(H_0\) and \(H_j\) be \(d\), i.e., they differ in \(d\) tag states.
	
	Note that each tag symbol spans $K$ RF source samples, with average source power $\sigma_s^2$ and noise variance $\sigma_u^2$. 
	Under linear detection in complex Gaussian noise, the probability that the detector confuses $H_0$ for $H_j$ depends on the Euclidean distance between their effective channel vectors, $\mathbf{w}_0$ and $\mathbf{w}_j$, relative to the noise level. The distinguishing distance for $H_0\to H_j$ 
	scales by $\sqrt{K\sigma_s^2}\|\mathbf{w}_0-\mathbf{w}_j\|$, compared against noise standard deviation $\sigma_u$. 	In a Gaussian detection framework \cite{jacobs1965principles}, the probability of misclassifying $H_0$ as $H_j$ is approximated by
	\begin{equation}
		P(H_0\to H_j )\approx Q\left( \sqrt{\frac{K \sigma_s^2}{\sigma_u^2}} \|\mathbf{w}_0-\mathbf{w}_j\|  \right),
	\end{equation}
	where $Q(x)=\tfrac{1}{\sqrt{2\pi}}\int_{x}^{\infty}e^{-t^2/2}dt$. 
	
	\noindent Putting these steps together yields a PEP of the form
	\begin{equation}
		\label{eq:PEP}
		P(H_0 \rightarrow H_j) \approx Q \left(\sqrt{\frac{K\sigma_s^2}{\sigma_u^2}\Delta(d)}\right),
	\end{equation}
	where $\Delta(d) = \|\mathbf{w}_0-\mathbf{w}_j\|^2$ is the squared Euclidean separation. Assuming that each tag contributes equally to the effective channel difference ($\Delta_0$), we can write $\Delta(d) \approx d\Delta_0$.
	
	The overall error probability can be upper bounded using the union bound. Grouping the error events according to the Hamming distance \(d\) (from $1$ to \(N\)) between the true and the erroneous hypotheses, we obtain
	\begin{equation}
		\label{eq:union_bound}
		P_e \leq \sum_{d=1}^{N} \binom{N}{d}  Q \left(\sqrt{\frac{K\sigma_s^2}{\sigma_u^2} d \Delta_0}\right),
	\end{equation}
	where \(\binom{N}{d}\) denotes the number of hypotheses that differ from the true hypothesis in exactly \(d\) tag states. To further simplify (\ref{eq:union_bound}), we can use a standard upper bound on the Q-function by employing the Chernoff bound, $Q(x) \leq \frac{1}{2}\exp(-\frac{x^2}{2})$, i.e.,
	\begin{equation}
		\label{eq:bound}
		P_e \leq \frac{1}{2}\sum_{d=1}^{N} \binom{N}{d} \exp \left(-\frac{K\sigma_s^2}{2\sigma_u^2}  d\Delta_0 \right).
	\end{equation}
	This bound is tractable because the summation is only over \(d = 1,2,\ldots,N\). This error probability depends on the number of tags \(N\), the number of samples \(K\), the signal power, and the effective per tag channel difference \(\Delta_0\).
	
	\subsection{Energy Detector}
	Energy detection (ED) provides a simple energy-based baseline. We develop an energy detector for our multi-tag AmBC system and derive a tractable decision rule based on the received signal energy.
	
	The ambient RF source signal \( s_k^{(t)} \) has a power of $\sigma_s^2$, i.e., $ \mathbb{E}[|s_k^{(t)}|^2] = \sigma_s^2$. The average energy of the received signal, computed over \(K\) samples, is  $E^{(t)} = \frac{1}{K}\sum_{k=1}^{K}\|\mathbf{x}_k^{(t)}\|^2$.

	\noindent Expanding the squared norm using $\mathbf{x}_k^{(t)} = \mathbf{w}_j s_k^{(t)}  + \mathbf{u}_k^{(t)}$, i.e.,
	\begin{equation*}
		E^{(t)} =  \frac{1}{K}\sum_{k=1}^{K} \Bigl( |s_k^{(t)}|^2 \|\mathbf{w}_j\|^2 + 2\operatorname{Re}( s_k^{(t)} \mathbf{w}_j^H \mathbf{u}_k^{(t)} ) + \|\mathbf{u}_k^{(t)}\|^2 \Bigr).
	\end{equation*}
	Since \(s_k^{(t)}\) and \(\mathbf{u}_k^{(t)}\) are independent and the noise \(\mathbf{u}_k^{(t)}\) is zero mean, the cross-term $ 2\operatorname{Re} (s_k^{(t)} \mathbf{w}_j^H \mathbf{u}_k^{(t)})$ has zero mean. By the law of large numbers, when averaged over a large number of samples \(K\) the contribution of this term will vanish.
	
	The test statistic $E^{(t)}$ is a sample average over \(K\) independent observations. By the central limit theorem, for large \(K\), \(E^{(t)}\) can be approximated as a Gaussian random variable under hypothesis \(H_j\) as $ E^{(t)} \sim \mathcal{N}(\delta_j,\gamma_j^2)$. The mean is
	\begin{equation}
		\label{eq:mean_E}
		\delta_j = \mathbb{E}\bigl[E^{(t)}\bigr] = \mathbb{E}[\|\mathbf{x}_k^{(t)}\|^2] = \sigma_s^2\|\mathbf{w}_j\|^2 + M \sigma_u^2 .
	\end{equation}
	Similarly, the variance of \(\|\mathbf{x}_k^{(t)}\|^2\) is given by
	\[
	\mathrm{Var}(\|\mathbf{x}_k^{(t)}\|^2) = \sigma_s^4 \|\mathbf{w}_j\|^4 + 2 \sigma_s^2\sigma_u^2\|\mathbf{w}_j\|^2 + M \sigma_u^4.
	\]
	Since \(E^{(t)}\) is the average over \(K\) samples, its variance is
	\begin{equation*}
		\label{eq:var_E}
		\gamma_j^2 = \mathrm{Var}\bigl(E^{(t)}\bigr) = \frac{1}{K}\left( \sigma_s^4\|\mathbf{w}_j\|^4 + 2\sigma_s^2\sigma_u^2\|\mathbf{w}_j\|^2 + M\sigma_u^4 \right).
	\end{equation*}
	
	In a multi-tag scenario with \(2^N\) hypotheses, the optimum decision rule is to choose the hypothesis that maximizes the likelihood of the observed energy \(E^{(t)}\). Under the Gaussian model for \(E^{(t)}\) given hypothesis \(H_j\),
	\[
	p\bigl(E^{(t)}|H_j\bigr) = \frac{1}{\sqrt{2\pi}\gamma_j} \exp\left(-\frac{\bigl(E^{(t)}-\delta_j\bigr)^2}{2\gamma_j^2}\right).
	\]
	Thus, the optimal decision rule is
	\begin{equation}
		\label{eq:joint_decision_opt}
		{j}^* = \argmax_{j \in \{0,1,\ldots,2^N-1\}} p\bigl(E^{(t)}|H_j\bigr).
	\end{equation}
	
	\begin{figure*}[t]
		\centering
		\resizebox{.8\linewidth}{!}{
			\begin{tikzpicture}[
				>=stealth,
				thick,
				node distance=1.2cm,
				font=\small
				]
				\definecolor{pilotcolor}{RGB}{255,230,180}    
				\definecolor{methodAcolor}{RGB}{255,204,204}  
				\definecolor{methodBcolor}{RGB}{166,206,227}  
				\definecolor{protoColor}{RGB}{255,255,210}    
				\definecolor{corrColor}{RGB}{210,255,210}     
				\definecolor{lrtColor}{RGB}{255,229,204}      
				\definecolor{bitsColor}{RGB}{255,255,204}     
				
				\tikzstyle{block} = [
				draw,
				thick,
				rectangle,
				align=center,
				fill=gray!15,
				rounded corners=2pt,
				minimum width=2.6cm,
				minimum height=1.0cm
				]
				\tikzstyle{arrow} = [->, thick]
				
				\def\myXshift{2.2cm}  
				
				\node[block, fill=pilotcolor, text width=2.8cm] (pilotA)
				{
					\textbf{Pilot Portion} \\
					\small (Covariance Matrix)
				};
				
				\node[block, fill=methodAcolor, text width=3.0cm,
				right=\myXshift of pilotA] (methodA)
				{
					\textbf{Approach I:} \\
					\textbf{Prototype Classifier} \\
					\small(EmbedNet)
				};
				
				\node[block, fill=lrtColor, text width=2.8cm,
				right=\myXshift of methodA] (proto)
				{
					\textbf{Form Prototypes} \\
					\small(Average pilot embeddings)
				};
				
				\node[block, fill=corrColor, text width=2.8cm,
				right=\myXshift of proto] (dataA)
				{
					\textbf{Data Portion} \\
					\small ($T-P$ symbols)
				};
				
				\node[block, fill=bitsColor, text width=2.6cm,
				right=\myXshift of dataA] (bitsA)
				{
					\textbf{Bit Decisions} \\
					\small(Nearest prototype)
				};
				
				\node[block, fill=pilotcolor, text width=2.8cm,
				below=0.8cm of pilotA] (pilotB)
				{
					\textbf{Pilot Portion} \\
					\small (One-Hot Correlation)
				};
				
				\node[block, fill=methodBcolor, text width=3.0cm,
				right=\myXshift of pilotB] (methodB)
				{
					\textbf{Approach II:} \\
					\textbf{Channel Estimation} \\
					\small(ChanEstNet)
				};
				
				\node[block, fill=lrtColor, text width=2.8cm,
				right=\myXshift of methodB] (lrt)
				{
					\textbf{LRT Detector} \\
					\small($2^N$ combos with channel coefficients)
				};
				
				\node[block, fill=corrColor, text width=2.8cm,
				right=\myXshift of lrt] (dataB)
				{
					\textbf{Data Portion} \\
					\small($T-P$ symbols)
				};
				
				\node[block, fill=bitsColor, text width=2.6cm,
				right=\myXshift of dataB] (bitsB)
				{
					\textbf{Bit Decisions} \\
					\small(LRT output)
				};
				
				\draw[arrow] (pilotA.east) -- (methodA.west);
				\draw[arrow] (methodA.east) -- (proto.west);
				\draw[arrow] (proto.east) -- (dataA.west);
				\draw[arrow] (dataA.east) -- (bitsA.west);
				
				\draw[arrow] (pilotB.east) -- (methodB.west);
				\draw[arrow] (methodB.east) -- (lrt.west);
				\draw[arrow] (lrt.east) -- (dataB.west);
				\draw[arrow] (dataB.east) -- (bitsB.west);
				
			\end{tikzpicture}
		}
		\caption{Illustration of two distinct deep learning approaches for multi-tag AmBC detection. 
			\emph{Approach I} (top row) uses a prototype based classifier (EmbedNet) on pilot observations, forms class prototypes, and then decodes data symbols by nearest prototype decisions.
			\emph{Approach II} (bottom row) performs channel estimation (ChanEstNet) on one-hot pilot correlations, applies an LRT to the data portion, and finally resolves bits via multi-hypothesis detection.}
		\label{fig:figgeneral}
		\vspace{-0.6cm}
	\end{figure*}
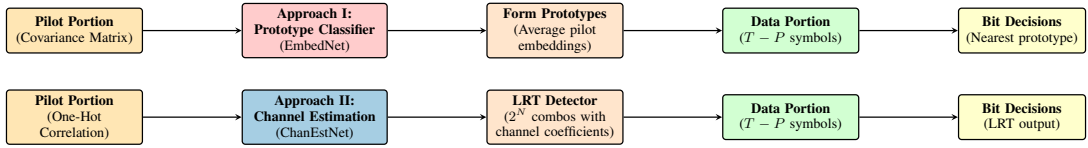

	Although likelihood based detectors are derived assuming perfect CSI, energy detection (ED) offers a simple alternative based on the received energy statistic. In AmBC, ED typically performs poorly when very weak backscatter components are masked by a strong direct path \cite{liu2013ambient}. At the same time, receivers that target near-LRT performance with conventional channel estimation require accurate and frequently updated CSI, which is difficult to obtain because of uncooperative ambient source and power constrained tags \cite{liu2020deep}.
		
	These limitations motivate data-driven deep learning based detection methods tailored to AmBC constraints. One approach, EmbedNet, bypasses explicit channel estimation and learns a direct mapping from covariance domain features of the received signal to the joint tag state using only a modest number of pilot symbols. The other, ChanEstNet, uses pilot symbols together to map correlation based features to effective channel coefficients that are then supplied to a classical LRT detector. Together, these two methods provide complementary CSI-free and CSI-aided deep learning solutions that relax the strict CSI requirements of classical detectors while respecting the practical constraints of multi-tag AmBC.

	\section{Deep Learning based Multi-Tag Detection Framework}
	In this section, we introduce two distinct deep learning approaches for detecting multiple tag symbols in our multi-tag AmBC system. Both approaches  address the challenges of signal processing and channel estimation in complex environments and rely on pilot-based information but differ in whether they directly classify bits or explicitly estimate channel coefficients before detection. Specifically,
	\begin{itemize}
		\item \textbf{Prototype Based Classification (EmbedNet):} 
		This first method learns to map pilot and data observations directly to class labels (tag states) without performing channel estimation. We employ a prototypical network framework where it produces an embedding for each received pilot signal, and we form prototypes by averaging embeddings of the same tag combination. New data signals are embedded and classified by their nearest prototype in the embedding space. This approach is purely discriminative and can adapt quickly with minimal pilot overhead.
		
		\item \textbf{Channel Estimation with LRT (ChanEstNet):}
		The second method uses a specialized CNN, to regress from a one-hot pilot correlation estimate to explicit channel coefficients. By separating out each tag in specific pilot periods, we obtain a set of naive correlation vectors and feed them into the network. Once we have the estimated channel, a standard likelihood ratio test can detect the multi-tag bits in the data portion. This approach integrates with classical detection theory, offering interpretability at the expense of requiring more structured pilot scheme.
	\end{itemize}
	
	These two solutions illustrate how deep learning can be leveraged for multi-tag backscatter detection in AmBC systems. Fig. \ref{fig:figgeneral} depicts the main processing blocks of EmbedNet and ChanEstNet, from pilot processing through detection of the data symbols. EmbedNet acts as a purely end-to-end detector that maps covariance based features directly to the joint tag state and is most appropriate when the primary objective is reliable detection without explicit channel estimates. ChanEstNet, in contrast, is intended for scenarios where the receiver requires both reliable detection and explicit channel information. It produces effective channel coefficients that are fed to the LRT detector and can also support higher layer functions, e.g., power control, link adaptation, and interference management through interpretable and model based metrics.
	
	From an architectural standpoint, both proposed methods employ compact convolutional networks rather than large fully connected or recurrent models. The input representations can be organized as small two dimensional arrays that encode spatial and pilot domain correlations, and convolutional filters are well suited to exploit this localized structure with relatively few parameters. This choice preserves the second order statistics of the received signal while avoiding the high parameter counts and memory requirements without ignoring the underlying geometry. Furthermore, the networks are intentionally kept shallow so that their computational cost can be kept minimal, which is critical for practical AmBC receivers\footnote{A quantitative comparison of the resulting complexity and inference time against classical LRT and ED baselines is provided in Section V.}.

	\subsection{Prototypical Network for Covariance Based Multi-Tag Detection}
	We present a prototypical network for multi-tag AmBC detection that learns from covariance matrix inputs without requiring explicit channel estimates. By using  pilot and data symbols as a few-shot classification task over each frame, we effectively handle $2^N$ multi-tag states while using only $P$ pilot samples for adaptation. Unlike direct supervised methods, prototypical networks generate prototypes (averaged embeddings) of each class from the pilot portion, then classify data symbols via simple distance comparisons in a learned embedding space. This approach is robust to diverse channel conditions, requires no additional parameter updates per frame, and efficiently accommodates the exponential growth of states with minimal overhead.

	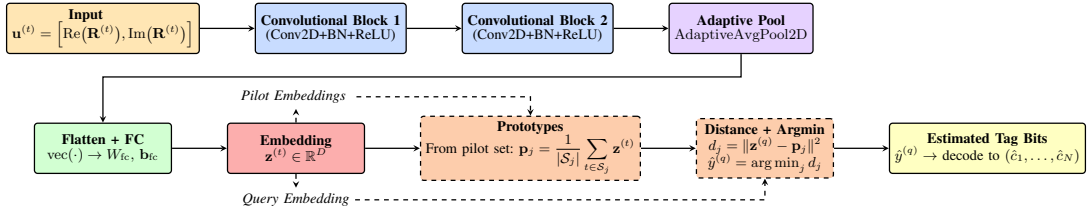
\begin{figure*}[t]
		\centering
		\resizebox{.8\linewidth}{!}{
			\begin{tikzpicture}[>=stealth, 
				font=\small, 
				block/.style={draw,thick,rectangle,rounded corners=2pt,minimum width=2.2cm,minimum height=1.1cm,align=center},
				sideblock/.style={draw,thick,dashed,rectangle,rounded corners=2pt,minimum width=2.3cm,minimum height=1.0cm,align=center,fill=gray!8},
				lineblock/.style={draw,thick,rectangle,rounded corners=2pt,minimum width=2.0cm,minimum height=0.8cm,align=center,fill=gray!15},
				skipcol/.style={rgb,255:red,160;green,160;blue,160},
				node distance=1.6cm]
				
				\definecolor{mycyan}{RGB}{220,255,255}
				\definecolor{myblue}{RGB}{200,220,255}
				\definecolor{mygreen}{RGB}{210,255,210}
				\definecolor{myorange}{RGB}{255,230,180}
				\definecolor{mypool}{RGB}{230,210,255}
				\definecolor{myembed}{RGB}{251,154,153}
				\definecolor{myest}{RGB}{255,255,153}
				\definecolor{myproto}{RGB}{253,205,172}
				\node[block, align=left, fill=myorange, minimum width=3.0cm] (input)
				{
					\textbf{ \hspace{1cm} Input} \\
					\small{$\mathbf{u}^{(t)}=\Bigl[\operatorname{Re}\bigl(\mathbf{R}^{(t)}\bigr), \operatorname{Im}\bigl(\mathbf{R}^{(t)}\bigr)\Bigr]$}
				};
				
				\node[block, fill=myblue, right=1.2cm of input, minimum width=3.0cm] (res1) {
					\textbf{Convolutional Block 1}\\[-1pt]
					\small(Conv2D+BN+ReLU)
				};
				
				\node[block, fill=myblue, right=1.2cm of res1, minimum width=3.0cm] (res2) {
					\textbf{Convolutional Block 2}\\[-1pt]
					\small(Conv2D+BN+ReLU)
				};
				
				\node[block, fill=mypool, right=1.2cm of res2, minimum width=3.0cm] (adapool) {
					\textbf{Adaptive Pool}\\[-1pt]
					\small$\mathrm{AdaptiveAvgPool2D}$
				};
				
				\node[block, fill=mygreen, below=1.5cm of input, minimum width=3.0cm] (fc) {
					\textbf{Flatten + FC}\\[-1pt]
					\small{$\mathrm{vec}(\cdot)\to W_{\mathrm{fc}},\,\mathbf{b}_{\mathrm{fc}}$}
				};
				
				\node[block, fill=myembed!80, right=1.2cm of fc, minimum width=3.0cm] (embedding) {
					\textbf{Embedding}\\
					\small{$\mathbf{z}^{(t)} \in \mathbb{R}^D$}
				};
				
				\node[sideblock, fill=myproto, right=1.2cm of embedding, minimum width=3.0cm] (protoBlock) {
					\textbf{Prototypes}\\
					\small From pilot set: $\mathbf{p}_j = \displaystyle\frac{1}{|\mathcal{S}_j|}\sum_{t \in \mathcal{S}_j}\mathbf{z}^{(t)}$
				};
				
				\node[sideblock, fill=myproto, right=1.2cm of protoBlock, minimum width=3.0cm] (distBlock) {
					\textbf{Distance + Argmin}\\
					\small $d_j = \|\mathbf{z}^{(q)} - \mathbf{p}_j\|^2$\\
					\small $\hat{y}^{(q)} = \argmin_{j} d_j$
				};
				
				\node[block, fill=myest!60, right=1.2cm of distBlock] (bitsOut) {
					\textbf{Estimated Tag Bits}\\
					\small $\hat{y}^{(q)}\to$ decode to $(\hat{c}_1,\dots,\hat{c}_N)$
				};
				
				\draw[->,thick] (input.east) -- (res1.west);
				\draw[->,thick] (res1.east) -- (res2.west);
				\draw[->,thick] (res2.east) -- (adapool.west);
				
				\draw[->,thick] (adapool.south) -- ++(0,-0.5) -| (fc.north);
				
				\draw[->,thick] (fc.east) -- (embedding.west);
				\draw[->,thick] (embedding.east) -- (protoBlock.west);
				\draw[->,thick] (protoBlock.east) -- (distBlock.west);
				\draw[->,thick] (distBlock.east) -- (bitsOut.west);

				\node[above=0.3cm of embedding, align=center] (pilotNote) { \emph{Pilot Embeddings} };
				\draw[dashed,->,thick] (embedding.north) -- (pilotNote.south);
				\draw[dashed,->,thick] (pilotNote.east)  -| (protoBlock.north);

				\node[below=0.3cm of embedding, align=center] (queryNote) {\emph{Query Embedding}};
				\draw[dashed,->,thick] (embedding.south) -- (queryNote.north);
				\draw[dashed,->,thick] (queryNote.east)  -| (distBlock.south);

			\end{tikzpicture}
		}
		\caption{Architecture of the proposed EmbedNet prototypical network for multi-tag AmBC detection. Covariance based inputs are processed by two convolutional blocks, adaptive pooling, and a fully connected layer to produce embeddings, which are classified via pilot derived nearest prototypes.}
		\label{fig:prototype}
		\vspace{-0.5cm}
	\end{figure*}

	For the $t$-th tag symbol out of $T$ total in a frame, the reader collects the observation matrix as
	\[
	\mathbf{X}^{(t)} = \Bigl[\mathbf{x}_1^{(t)},\mathbf{x}_2^{(t)},\dots,\mathbf{x}_K^{(t)}\Bigr]\in\mathbb{C}^{M\times K},
	\]
	where $\mathbf{x}_k^{(t)} \in \mathbb{C}^{M}$ is the $k$-th instantaneous snapshot across $M$ antennas.
		
	While raw observation matrix can be used as an input to the EmbedNet, it strongly depends on phases and the unknown ambient waveform $s_k^{(t)}$. On the other hand, the covariance matrix, $\mathbf{R}^{(t)}$, partially averages out unknown phases across $K$ samples, stabilizing the representation and achieving better detection performance. This leads to improved generalization since the channel or ambient source modulations vary. From \eqref{eq:rx_signal}, for a joint tag state during symbol period $t$, we have $\mathbf{w}^{(t)}=\mathbf{h}+\sum_{i=1}^N \alpha_i f_i \mathbf{g}_i c_i^{(t)}$, so that $\mathbf{x}_k^{(t)}=\mathbf{w}^{(t)} s_k^{(t)}+\mathbf{u}_k^{(t)}$. Then $\mathbb{E}\!\left[\mathbf{R}^{(t)} \mid \mathbf{w}^{(t)}\right]=\sigma_s^2 \mathbf{w}^{(t)}(\mathbf{w}^{(t)})^\mathsf{H}+\sigma_u^2\mathbf{I}_M$, which motivates the use of covariance features as modulation-agnostic signatures of the joint tag state. Hence, we compute a sample covariance matrix as
	\begin{equation}
		\label{eq:cov_reminder}
		\mathbf{R}^{(t)} = \frac{1}{K}\mathbf{X}^{(t)}\bigl(\mathbf{X}^{(t)}\bigr)^\mathsf{H} \in \mathbb{C}^{M\times M},
	\end{equation}
	which combines the direct plus backscatter signals in a second order statistic. To feed real-valued inputs to the neural network, we concatenate the real and imaginary parts of \(\mathbf{R}^{(t)}\), i.e.,
	\begin{equation}
		\label{eq:input_tensor}
		\mathbf{u}^{(t)}=\
		\Bigl[\operatorname{Re}\bigl(\mathbf{R}^{(t)}\bigr), \operatorname{Im} \bigl(\mathbf{R}^{(t)}\bigr)
		\Bigr] \in \mathbb{R}^{2 \times M \times M}.
	\end{equation}
	
	For each sample $\mathbf{u}^{(t)}$\footnote{For a system with $M=4$ receive antennas, the input dimensions are $2 \times 4 \times 4$.}, let the corresponding tag state for symbol \(t\) be given as $y^{(t)} \in \bigl\{0,1,\dots,2^N - 1\bigr\}$.

	For a single frame with $T$ symbol periods, we collect a sample training set as
	\begin{equation}
		\mathcal{D} =
		\Bigl\{
		\bigl(\mathbf{u}^{(1)},y^{(1)}\bigr),
		\bigl(\mathbf{u}^{(2)},y^{(2)}\bigr),\dots,
		\bigl(\mathbf{u}^{(T)},y^{(T)}\bigr)
		\Bigr\}.
	\end{equation}
	
	We note that each frame consists of $T$ tag symbols which are split as follows: 1) $P$ pilot symbols, which are known to the reader for each frame. 2) $T-P$ data symbols, which carry the actual unknown multi-tag bits that will be detected. Following the frame structure, we divide the dataset into two parts, i.e.,
	\begin{enumerate}
		\item \emph{Pilot (Support) Subset}: The first $P$ symbol periods, 
		\[
		\mathcal{D}^{p}  = \Bigl\{ \bigl(\mathbf{u}^{(t)}, y^{(t)}\bigr) \mid t=1,\dots,P \Bigr\},
		\]
		where these pilot symbols serve as labeled support examples for the prototypical network.
		\item \emph{Data (Query) Subset}: The remaining $T-P$ symbol periods,
		\[
		\mathcal{D}^{q}  = \Bigl\{ \bigl(\mathbf{u}^{(t)}, y^{(t)}\bigr) \mid t=P+1,\dots,T \Bigr\},
		\]
		which contain the \emph{unknown} tag states in a real deployment. At test time, these are unlabeled and must be predicted by the trained model.
	\end{enumerate}

	A prototypical network deals with classification when each class has few labeled samples (the support set). It is composed of two main parts, i.e.,
	\begin{itemize}
		\item \emph{Embedding function}: A deep neural network, denoted by $\mathrm{netA}_{\boldsymbol{\theta}}(\cdot)$ with learnable parameters $\boldsymbol{\theta}$, transforms each input  into a feature vector (embedding) in $\mathbb{R}^D$.
		\item \emph{Prototype based classification}: Given the $P$ pilot embeddings grouped by their label, we compute prototype vectors. Then, each unlabeled data embedding is classified by the prototype  that is closest in Euclidean distance.
	\end{itemize}

	We first present the neural network architecture used for EmbedNet, illustrated in Fig. \ref{fig:prototype} and detailed below.
	
	\paragraph{Convolutional Layers}
	A two dimensional convolution (\textsc{Conv2D}) layer applies a trainable kernel across two spatial dimensions to extract local features. Mathematically, for an input $\mathbf{x}\in\mathbb{R}^{c\times h\times w}$ (with $c$ channels and spatial size $h\times w$) and a set of learnable parameters $\boldsymbol{\omega}$, a single convolutional output channel can be expressed as
	\begin{equation}
		\mathrm{Conv2D}(\mathbf{x};\boldsymbol{\omega})[i,j] = \sum_{u,v}\sum_{r=1}^{c} \boldsymbol{\omega}_{r,u,v}\mathbf{x}[r,i-u,j-v],
		\label{eq:conv2d_math}
	\end{equation}
	where the sums over $(u,v)$ index the kernel’s spatial extent, and $r$ indexes the input channels.
	
	\paragraph{Batch Normalization (BN)}
	Batch normalization is often applied to stabilize training which normalizes activation of each channel by subtracting a batch wise mean and dividing by a standard deviation, then scales and shifts them with learnable parameters. Mathematically, if $z$ is an activation in the channel, BN transforms it as
	\begin{equation}
		\mathrm{BN}(z)= \gamma \frac{z - \mu_{\mathrm{batch}}}{\sigma_{\mathrm{batch}} + \epsilon} +\beta,
	\end{equation}
	where $\mu_{\mathrm{batch}}$, $\sigma_{\mathrm{batch}}$ are the mean and standard deviation over the mini-batch, $\gamma$ and $\beta$ are learnable scale and shift, and $\epsilon$ is a small constant to avoid division by zero.
	
	\paragraph{ReLU Activation}
	A Rectified Linear Unit (ReLU) is a pointwise non-linearity function that accelerates convergence and helps alleviate the vanishing gradient problem in deeper networks and for an input $x$ defined as
	\[
	\mathrm{ReLU}(x) =\max\{0,x\}.
	\] 
	
	\subsubsection{Convolutional Block}
	We group a Conv2D layer, batch normalization, and a ReLU activation into a convolutional block and apply two such blocks in sequence. Given an input covariance feature tensor $\mathbf{u}^{(t)}\in\mathbb{R}^{2\times M\times M}$, the successive convolutional blocks compute
	\begin{align*}
		\mathbf{z}_1^{(t)}
		&=
		\mathrm{BN}\Bigl(\mathrm{Conv2D}\bigl(\mathbf{u}^{(t)}\bigr)\Bigr),
		\quad
		\mathbf{y}_1^{(t)}
		=
		\mathrm{ReLU}\bigl(\mathbf{z}_1^{(t)}\bigr),
		\\[6pt]
		\mathbf{z}_2^{(t)}
		&=
		\mathrm{BN}\Bigl(\mathrm{Conv2D}\bigl(\mathbf{y}_1^{(t)}\bigr)\Bigr),
		\quad
		\mathbf{y}_2^{(t)}
		=
		\mathrm{ReLU}\bigl(\mathbf{z}_2^{(t)}\bigr),
	\end{align*}
	where $\mathbf{y}_2^{(t)}$ denotes the final convolutional feature map that is provided to the subsequent adaptive pooling stage.
	
	\subsubsection{Adaptive Pooling}
	Following the convolutional blocks, we reduce the spatial dimensions of the feature map $\mathbf{y}_2^{(t)}$ to fixed lower dimensions $(p,q)$ using an adaptive average pooling operator as
	\begin{equation}
		\mathbf{y}_\mathrm{pool}^{(t)} = \mathrm{AdaptiveAvgPool2D}\bigl(\mathbf{y}_2^{(t)}\bigr),
	\end{equation}
	which partitions the feature map into smaller regions and averages each region to produce a uniform size output.

	\subsubsection{Fully Connected Output}
	Finally, we flatten $\mathbf{y}_\mathrm{pool}^{(t)}$ into a vector $(\mathbf{y}_\mathrm{pool}^{(t)})^{\mathrm{vec}}$ and apply a fully connected layer to output a $D$-dimensional embedding vector, i.e.,
	\begin{equation}
		\mathbf{z}^{(t)}=\mathbf{w}_{\mathrm{fc}} \bigl(\mathbf{y}_\mathrm{pool}^{(t)}\bigr)^{\mathrm{vec}} + \mathbf{b}_{\mathrm{fc}},
	\end{equation}
	where $\mathbf{w}_{\mathrm{fc}}$ and $\mathbf{b}_{\mathrm{fc}}$ are trainable parameters, and $\mathbf{z}^{(t)}\in\mathbb{R}^D$ represents the feature embedding of the $t$-th input. The complete architecture parameters of EmbedNet are summarized in Table \ref{tab:embednet}.
	
	\begin{table}[t]
	\centering
	\caption{\small EmbedNet architecture parameters}
	\label{tab:embednet}
	
	\renewcommand{\arraystretch}{1.2}
	\resizebox{\columnwidth}{!}{
	\begin{tabular}{lccccc}
	\hline
	\textbf{Layer} & \textbf{Type} & \textbf{Kernel} & \textbf{Filters/Units} & \textbf{Output Shape} & \textbf{Activation} \\
	\hline
	Input & -- & -- & -- & $2 \times M \times M$ & -- \\
	Conv2D-1 & Conv2D + BN & $3 \times 3$ & 32 & $32 \times M \times M$ & ReLU \\
	Conv2D-2 & Conv2D + BN & $3 \times 3$ & 64 & $64 \times M \times M$ & ReLU \\
	Adaptive Pool & Avg Pool 2D & -- & -- & $64 \times 4 \times 4$ & -- \\
	Embedding (FC) & Linear & -- & 64 & 64 & -- \\
	Classification & Prototype Dist. & -- & $2^N$ classes & $2^N$ & Softmax \\
	\hline
	\end{tabular}}
    \vspace{-0.5cm}
	\end{table}
	
	We now perform classification of tags' bits by measuring distances between the embeddings and prototypes. Specifically, we first define class indexes for each class and denote it by $j\in\{0,1,\dots,2^N-1\}$. In other words, class indexes represent different hypotheses which corresponds to the classes in our approach. We then define the set of pilot indices whose label is $j$ as  
	$\mathcal{S}_j = \{t \mid 1 \le t \le P,y^{(t)} = j\}$.
	
	Following that, we define the prototype for class $j$ as the average embedding of all pilot examples $t$ whose label is $j$, i.e.,
	\begin{equation}
		\mathbf{p}_j=
		\frac{1}{|\mathcal{S}_j|}\sum_{t\in \mathcal{S}_j} \mathrm{netA}_{\boldsymbol{\theta}}\Bigl(\mathbf{u}^{(t)}\Bigr).
		\label{eq:proto_equation_long} 
	\end{equation}

	Next, for each unlabeled data symbol index $q \in\{P+1,\dots,T\}$ from the data subset $\mathcal{D}^{q}$, we obtain embedding of each data symbol via the trained network, i.e., $\mathbf{z}^{(q)} = \mathrm{netA}_{\boldsymbol{\theta}}\bigl(\mathbf{u}^{(q)}\bigr)$. We then compute the squared distance to each prototype $\mathbf{p}_j$ as
	\begin{equation}
		d_j^{(q)} =\|\mathbf{z}^{(q)}-\mathbf{p}_j\|^{2}.
	\end{equation}
	As a final step, we assign the predicted class as
	\begin{equation}
		\hat{y}^{(q)} =\argmin_{j}d_j^{(q)},
	\end{equation}
	which picks the class whose prototype in the embedding space is nearest to $\mathbf{z}^{(q)}$. This decision rule requires no further parameter updates and only involves computing $2^N$ distances per query sample (data symbol). This nearest prototype rule is extremely lightweight at test time and adapts to each new frame simply by re-computing prototypes from the pilot portion of that frame. Thus, the same neural network $\mathrm{netA}_{\boldsymbol{\theta}}(\cdot)$ can flexibly adapt to environment of each frame with minimal overhead, leveraging only $P$ pilot examples.
	
	\begin{figure*}[t]
		\centering
		\resizebox{\linewidth}{!}{
			\begin{tikzpicture}[>=stealth, font=\small, thick, node distance=1.6cm]
				
				\tikzstyle{block} = [draw, thick, rectangle, align=center, fill=gray!15, rounded corners=2pt, minimum width=2.2cm, minimum height=1.1cm]
				\tikzstyle{arrow} = [->, thick]

				\definecolor{mycyan}{RGB}{220,255,255}
				\definecolor{myblue}{RGB}{200,220,255}
				\definecolor{mygreen}{RGB}{210,255,210}
				\definecolor{myorange}{RGB}{255,230,180}
				\definecolor{mypool}{RGB}{230,210,255}
				
				\node[block, fill=myorange] (inp) {
					\begin{tabular}{c}
						\textbf{Input} \\[-2pt]
						$\mathbf{c}_n = \left[\operatorname{Re}(\widehat{\mathbf{w}}^{(i)}_{\mathrm{corr}}), \operatorname{Im}(\widehat{\mathbf{w}}^{(i)}_{\mathrm{corr}}) \right]$
					\end{tabular}
				};
				
				\node[block, fill=myblue, right=1cm of inp] (conv1) {
					\textbf{Convolutional Block 1}\\[-1pt]
					\small(Conv2D+BN+ReLU)
				};
				
				\node[block, fill=myblue, right=1cm of conv1] (conv2) {
					\textbf{Convolutional Block 2}\\[-1pt]
					\small(Conv2D+BN+ReLU) 
				};
				
				\node[block, fill=myblue, right=1cm of conv2] (conv3) {
					\textbf{Convolutional Block 3}\\[-1pt]
					\small(Conv2D+BN+ReLU)
				};
				
				\node[block, fill=mypool, below=1cm of conv3] (pool) {
					\textbf{Adaptive Pool}\\[-1pt]
					\small$\mathrm{AdaptiveAvgPool2D}$
				};
				
				\node[block, fill=mygreen, right=1cm of pool] (fc) {
					\textbf{Flatten + FC}\\[-1pt]
					\small{$\mathrm{vec}(\cdot)\to W_{\mathrm{fc}},\mathbf{b}_{\mathrm{fc}}$}
				};
				
				\node[block, fill=myorange, right=1cm of fc] (out) {
					\begin{tabular}{c}
						\textbf{ChanEstNet Output} \\
						$\bigl[\widehat{\mathbf{h}},\widehat{\mathbf{v}}_{1},\dots,\widehat{\mathbf{v}}_{N}\bigr]$
					\end{tabular}
				};
				
				\node[block, fill=mycyan, right=1cm of out, minimum width=3.1cm] (lrt) {
					\textbf{LRT Detector} \\
					$\hat{c}_{1},\ldots,\hat{c}_{N}$ \\
					\textbf{Detected Bits}
				};
				
				\draw[arrow] (inp) -- (conv1);
				\draw[arrow] (conv1) -- (conv2);
				\draw[arrow] (conv2) -- (conv3);
				\draw[arrow] (conv3) -- (pool);
				\draw[arrow] (pool.east) -- (fc.west);
				\draw[arrow] (fc.east) -- (out.west);
				\draw[arrow] (out.east) -- (lrt.west);

			\end{tikzpicture}
		}
		\caption{Architecture of the proposed ChanEstNet channel estimation network for multi-tag AmBC. One-hot pilot correlations are processed by convolutional blocks, adaptive pooling, and a fully connected layer to produce channel estimates, which are then used by an LRT detector for final bit decisions.}
		\label{fig:onehot_cnn_arch_lrt}
		\vspace{-0.5cm}
	\end{figure*}
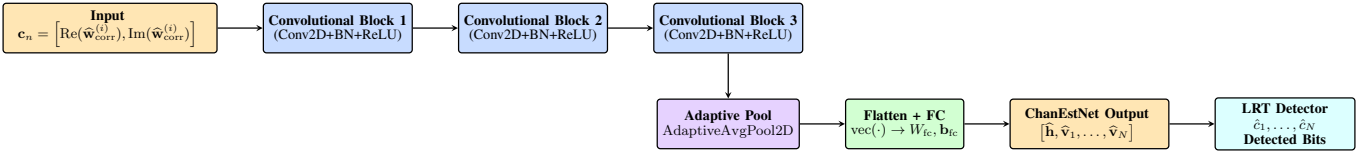

	We adopt the standard episodic approach used in prototypical networks \cite{snell2017prototypical} to train $\mathrm{netA}_{\boldsymbol{\theta}}(\cdot)$, which is given as
	\begin{itemize}
		\item We define a single task by randomly generating a frame of length $T$. This includes $P$ pilot symbols, each with a known class in $\{0,\dots,2^N-1\}$, and $T-P$ unlabeled data. We define a support set $\mathcal{D}^{p}$ and a query set $\mathcal{D}^{q}$.
		\item On the support set, we compute embeddings for all $P$ pilot covariances. Then, for each class $j\in\{0,\dots,2^N-1\}$, we average all embeddings that belong to $j$, forming $\mathbf{p}_j$ as in \eqref{eq:proto_equation_long}. Next, for each query sample $q\in\{P+1,\dots,T\}$, we compute a vector of distances $d_j^{(q)} $ as logits. A standard cross-entropy function against the true query label $y^{(q)}$ produces a task loss, i.e., 
		\begin{equation}
			\ell_\mathrm{task}(\theta) =\frac{1}{Q}\!\!\!\sum_{q=P+1}^{T}
			\!\!\!\!\!-\ln \left[
			\frac{\exp(-\|\mathbf{z}^{(q)}-\mathbf{p}_{y^{(q)}}\|^2)}
			{\sum_{j=0}^{2^N-1}\exp(-\|\mathbf{z}^{(q)}-\mathbf{p}_j\|^2)}
			\right], 
		\end{equation}
		where $Q = T-P$ and using the loss function, we perform gradient-based updates on $\boldsymbol{\theta}$ using Adam optimizer \cite{kingma2014adam}.
		\item We gather different frames under diverse random channels $(\mathbf{h}, f_i, \mathbf{g}_i)$, varied SNR, and random pilot states for $\mathrm{netA}_{\boldsymbol{\theta}}(\cdot)$ to generalize well. Over thousands of frames, the network learns channel invariant features that separate the multi-tag classes effectively. This is crucial for real-time deployment, since we never know which channels or reflection coefficients will appear.
	\end{itemize}
	
	\subsection{One-Hot Pilot Correlation for Channel Estimation}
	We now propose a one-hot pilot correlation approach combined with a convolutional neural network to estimate channel coefficients in a multi-tag AmBC system. Unlike the direct classification or prototype based methods, this framework explicitly learns an estimate of the channel by exploiting the orthogonal pilot structure of the pilot and improving a naive correlation estimate using a deep neural network. The resulting channel estimates can then be used in a likelihood ratio test. 
	
	We first explain how we design the pilot symbols for channel estimation. We note that a single frame has $T$ tag symbol periods, out of which $P$ are used for pilots and the remaining ($T-P$) for data. For ChanEstNet, we allocate the pilot portion to repeated one-hot blocks that isolate the direct path and each tag backscatter component. Specifically, each block uses one of the $(N+1)$ configurations:
	\begin{itemize}
		\item All-Off Pilot Symbol: During this pilot symbol period, all $N$ tags remain in the non-reflective state ($c_i^{(t)}=0$). Hence the effective channel is just the direct path, $\mathbf{h}$.
		\item Per-Tag Pilot Symbol $i$: In the $i$-th one-hot pilot symbol period, only tag $i$ is turned on ($c_i^{(t)}=1$),
		while all other tags remain off. Thus the effective channel becomes
		$\mathbf{h} + \mathbf{v}_i,$
		where $\mathbf{v}_i = \alpha_i f_i \mathbf{g}_i.$
	\end{itemize}
	
	We schedule the $P$ pilot symbols across these $(N+1)$ one-hot configurations so that configuration $i$ appears $P_i$ times, i.e., $\sum_{i=0}^N P_i = P$. We define $\mathcal{P}_i$ as the set of pilot indices assigned to configuration $i$. For each pilot symbol $t\in\{1,\dots,P\}$, we collect $K$ observations $\mathbf{x}_{k}^{(t)}$ and assume the corresponding ambient symbols $s_{k}^{(t)}$ are known or reliably estimated at the reader. For each configuration $i$, we compute a naive correlation for each associated pilot and average across its $P_i$ occurrences to obtain one correlation vector, i.e.,
	\begin{equation}
		\label{eq:corr_naive}
		\widehat{\mathbf{w}}^{(i)}_{\mathrm{corr}}= \frac{1}{P_i}\sum_{t\in\mathcal{P}_i}\frac{\sum_{k=1}^{K}\bigl[s_{k}^{(t)}\bigr]^{*}\mathbf{x}_{k}^{(t)}}{\sum_{k=1}^{K} \bigl|s_{k}^{(t)}\bigr|^2},
		\quad i=0,1,\dots,N.
	\end{equation}
	This operation cancels the random phases from $s_{k}^{(t)}$. As a result, $\widehat{\mathbf{w}}^{(i)}_{\mathrm{corr}}$ provides a correlation based estimate of the effective channel for configuration $i$, compressing the $K$ received samples into a single $M$-dimensional vector. These correlation vectors form a modulation-agnostic summary of the pilot frame and serve as the input features to ChanEstNet.

	We then stack these $(N+1)$ correlation outputs $\widehat{\mathbf{w}}^{(i)}_{\mathrm{corr}} \in \mathbb{C}^{M}$, into a real-valued 3D array $\mathbf{c}_n \in \mathbb{R}^{2 \times(N+1)\times M}$. Specifically, for each $i$,
	\begin{equation}
		\mathbf{c}_n\bigl[0,\,i,:\bigr] =\operatorname{Re}\bigl(\widehat{\mathbf{w}}_{\mathrm{corr}}^{(i)}\bigr),
		\quad \mathbf{c}_n\bigl[1,\,i,:\bigr] =\operatorname{Im}\bigl(\widehat{\mathbf{w}}_{\mathrm{corr}}^{(i)}\bigr).
	\end{equation}

	The label for each pilot symbol in our one-hot pilot dataset is the collection of true channel vectors \(\mathbf{h},\mathbf{v}_1,\dots,\mathbf{v}_N\) gathered into a real-valued matrix. We then assemble our ground-truth label matrix \(\mathbf{d}_n\) by stacking the real and imaginary parts of \(\mathbf{h}\) and each \(\mathbf{v}_i\), i.e.,
	\begin{equation}
		\label{eq:label_Gstar}
		\mathbf{d}_n =
		\begin{bmatrix}
			\left[\operatorname{Re}(\mathbf{h}), \operatorname{Im}(\mathbf{h}) \right] \\[4pt]
			\left[\operatorname{Re}(\mathbf{v}_1), \operatorname{Im}(\mathbf{v}_1) \right] \\
			\vdots \\[2pt]
			\left[\operatorname{Re}(\mathbf{v}_N), \operatorname{Im}(\mathbf{v}_N) \right]
		\end{bmatrix}
		\in\mathbb{R}^{(N+1)\times2M}.
	\end{equation}

	Hence, each row of \(\mathbf{d}_n\) corresponds to the real-valued representation of one complex channel vector (either \(\mathbf{h}\) or \(\mathbf{v}_i\)). Specifically, \(\mathbf{d}_n\) serves as the label associated with the correlation tensor in our dataset. For each frame, the $P$ pilot symbols are grouped by one-hot configuration and repeated over to produce a single training pair $(\mathbf{c}_n,\mathbf{d}_n)$, i.e.,
	\begin{equation}
		\mathcal{D}^{c} =\{(\mathbf{c}_n, \mathbf{d}_n)\}.
	\end{equation}

	We adopt a neural network with one-hot pilot correlation for channel estimation, denoted by $\mathrm{netB}_{\boldsymbol{\phi}}(\mathbf{c}_n):
	\mathbb{R}^{2\times(N+1)\times M}\rightarrow \mathbb{R}^{(N+1)\times 2M}$ with parameter set $\boldsymbol{\phi}$, as illustrated in Fig. \ref{fig:onehot_cnn_arch_lrt}.
	\begin{enumerate}
		\item \emph{Conv2D Blocks:} We stack three convolutional layers, each with an increasing number of output channels, followed by batch normalization and ReLU activation. Mathematically, for layer $\ell=1,2,3$, we have
		\begin{equation}
			\mathbf{z}_\ell^{(n)} = \mathrm{BN}_\ell\bigl(\mathrm{Conv2D}_\ell(\mathbf{y}_{\ell-1}^{(n)})\bigr),
			\quad \mathbf{y}_\ell^{(n)} = \mathrm{ReLU}\bigl(\mathbf{z}_\ell^{(n)}\bigr),
		\end{equation}
		where $\mathbf{y}_0^{(n)} = \mathbf{c}_n$ is the network input.
		\item \emph{Adaptive Pooling:} We apply $\mathrm{AdaptiveAvgPool2D}$ to compress the spatial dimension to obtain a uniform shape. The output from the final convolutional block, $\mathbf{y}_3^{(n)}$, is passed into an $\mathrm{AdaptiveAvgPool2D}$, producing
		\begin{equation}
			\mathbf{y}_{\mathrm{pool}}^{(n)}  = \mathrm{AdaptiveAvgPool2D}\bigl(\mathbf{y}_3^{(n)}\bigr).
		\end{equation}
		\item \emph{Flatten + Dense:} We flatten $\mathbf{y}_{\mathrm{pool}}^{(n)}$ into a vector $(\mathbf{y}_{\mathrm{pool}}^{(n)})^{\mathrm{vec}}$ and apply a fully connected layer to map it into real outputs, i.e.,
		\begin{equation}
			\mathbf{d}_{\mathrm{out}}^{(n)} = \mathbf{W}_{\mathrm{fc}}(\mathbf{y}_{\mathrm{pool}}^{(n)})^{\mathrm{vec}} +\mathbf{b}_{\mathrm{fc}}  
			\in \mathbb{R}^{(N+1) \times 2M}.
		\end{equation}
		where $\mathbf{W}_{\mathrm{fc}}$ and $\mathbf{b}_{\mathrm{fc}}$ are the network weights and bias of the fully connected layer. $\mathbf{d}_{\mathrm{out}}^{(n)}$ is output of the network which contains the channel estimates in a concatenated form of real and imaginary parts, i.e., $\mathbf{d}_{\mathrm{out}}^{(n)} = \mathrm{netB}_{\boldsymbol{\phi}}(\mathbf{c}_n)$.
	\end{enumerate}
	
	Finally, we reshape $\mathbf{d}_{\mathrm{out}}^{(n)} \in \mathbb{R}^{(N+1)\times 2M}$ back into a complex matrix to obtain channel estimates as
	\begin{equation}
		\label{eq:netB_output}
		\mathbf{d}_{\mathrm{out_2}}^{(n)} =
		\begin{bmatrix}
			\widehat{\mathbf{h}}\\[3pt]
			\widehat{\mathbf{v}}_1\\
			\vdots\\
			\widehat{\mathbf{v}}_N
		\end{bmatrix} \in\mathbb{C}^{(N+1)\times M},
	\end{equation}
	where each row has channel estimates in a complex vector. The complete architecture parameters of ChanEstNet are summarized in Table \ref{tab:chanestnet}.
	
	\begin{table}[t]
	\centering
	\caption{\small ChanEstNet architecture parameters}
	\label{tab:chanestnet}
	
	\renewcommand{\arraystretch}{1.2}
	\resizebox{\columnwidth}{!}{
	\begin{tabular}{lccccc}
	\hline
	\textbf{Layer} & \textbf{Type} & \textbf{Kernel} & \textbf{Filters/Units} & \textbf{Output Shape} & \textbf{Activation} \\
	\hline
	Input & -- & -- & -- & $2 \times (N\!+\!1) \times M$ & -- \\
	Conv2D-1 & Conv2D + BN & $3 \times 3$ & 32 & $32 \times (N\!+\!1) \times M$ & ReLU \\
	Conv2D-2 & Conv2D + BN & $3 \times 3$ & 64 & $64 \times (N\!+\!1) \times M$ & ReLU \\
	Conv2D-3 & Conv2D + BN & $3 \times 3$ & 128 & $128 \times (N\!+\!1) \times M$ & ReLU \\
	Adaptive Pool & Avg Pool 2D & -- & -- & $128 \times 4 \times 4$ & -- \\
	Embedding (FC) & Linear & -- & 256 & 256 & ReLU \\
	Output (FC) & Linear & -- & $(N\!+\!1)2M$ & $(N\!+\!1)2M$ & -- \\
	\hline
	\end{tabular}}
    \vspace{-0.5cm}
	\end{table}	
	
	Following channel estimation process through ChanEstNet, we use a classical multi-hypothesis LRT for data detection as described in section \ref{subsec:LRT}, i.e.,
	\begin{itemize}
		\item We form the \(2^N\) possible combined channels by using the estimates \(\widehat{\mathbf{h}}\) and each \(\widehat{\mathbf{v}}_{i}\), 
		\begin{equation*}
			\widehat{\mathbf{w}}_{j}=\widehat{\mathbf{h}} + \sum_{i=1}^{N} b_{j,i} \widehat{\mathbf{v}}_{i},
			\quad j=0,\dots,2^N-1.
		\end{equation*}
		\item For each data symbol consisting of \(K\) samples and unknown data bits, we select the hypothesis index as 
		\[
		j^{*}
		=\argmax_{j\in\{0,\dots,2^N-1\}}\sum_{k=1}^K \log p \Bigl(\mathbf{x}_{k}^{(\mathrm{data})}\big|\widehat{\mathbf{w}}_{j},s_{k}^{(\mathrm{data})}\Bigr),
		\]
		deciding on the index \(j^{*}\) which maximizes the likelihood function. The bits corresponding to \(j^{*}\) become the final multi-tag decisions.
	\end{itemize}

	We train the ChanEstNet network by minimizing a mean-squared error (MSE) between the network output and the ground-truth channel vectors which is defined as 
	\begin{equation}
		\label{eq:onehotcnn_mseloss_updated}
		\mathcal{L}_{\mathrm{MSE}}(\phi) =  \sum_{n=1}^{(N+1)} \| \mathbf{d}_{\mathrm{out}}^{(n)}- \mathbf{d}_n \|^{2},
	\end{equation}
	where the lost function is averaged over all pilot samples in a mini-batch. Minimizing the loss with respect to $\phi$ trains the network to make each channel estimate closely match the corresponding true channel vector $\mathbf{h}$ or $\mathbf{v}_i$, thus aligning the predictions with the ground-truth. After optimization by using Adam optimizer \cite{kingma2014adam}, $\mathrm{netB}_{\boldsymbol{\phi}}(\cdot)$ robustly transforms each naive correlation vector $\mathbf{c}_n$ into accurate channel estimates.

	The training continues until sufficient number of epochs is reached by producing new frames with random direct channel \(\mathbf{h}\in\mathbb{C}^M\) and backscatter paths \(\mathbf{v}_i = \alpha_i f_i \mathbf{g}_i\). After sufficient training, the ChanEstNet learns to map each raw correlation input, $\mathbf{c}_n$, into an accurate channel estimates $\widehat{\mathbf{h}}$ and $\widehat{\mathbf{v}}_i$ across varied channel and SNR conditions.
	
	After training is complete, during real-time deployment, the final step is data detection on the $T-P$ data symbols. Note that we do not feed data symbols into the ChanEstNet, we only rely on the channel estimates from the pilot set. For each data symbol, we perform LRT using estimates out of \(\mathrm{netB}_{\boldsymbol{\phi}}(\cdot)\). The final detected bits are the binary representation of $j^{*}$.

	\vspace{-0.1cm}
	\section{Numerical Results}
	In this section, we present extensive numerical results that compare the performance of our two proposed deep learning approaches for multi-tag AmBC detection against both theoretical benchmarks, including an optimal LRT with CSI and an energy detector (ED) baseline. Specifically, we consider
	\begin{itemize}
		\item EmbedNet: A discriminative method that uses a pilot-embedded network and forms prototypes of each tag state combination for final classification.
		\item ChanEstNet: A regression-based approach wherein a neural network estimates the channel coefficients via one-hot correlation, followed by a  multi-hypothesis LRT.
		\item Theoretical LRT with CSI: A baseline that assumes perfect channel knowledge and performs an optimal log-likelihood ratio test.
		\item ED baseline: An energy-based multi-hypothesis detector operating on the received energy statistic.
	\end{itemize}

	We examine an AmBC system with either two or three single antenna passive tags, a single antenna RF source, and a reader equipped with $M=4$ antennas. The average backscatter to direct link ratio is set to \(\zeta_i = -20\) dB for all tags, capturing typical low-power backscatter conditions. The channels for forward, direct, and backscatter channels are modeled as Rayleigh fading, i.e., \(\mathbf{h}\sim\mathcal{CN}(\mathbf{0}, \mathbf{I}_M)\), \(\mathbf{g}_i\sim\mathcal{CN}(\mathbf{0}, \mathbf{I}_M)\), and \(f_i\sim\mathcal{CN}(0,1)\), ensuring that all links are independently scattered. This channel model captures the typical multipath propagation in AmBC systems where low-power backscattered signals undergo significant scattering without a dominant line of sight path, which is widely accepted in the literature \cite{liu2021deep, Yang2023Non-Coherent}. We set $T=160$ and $P=32$ as a representative frame design that keeps pilot overhead moderate while supporting scalability up to $5$ tags (i.e., $2^N=32$ joint states). The remaining $T-P=128$ data symbols per frame provide a sufficiently long block for stable BER estimation.
	
	For the two proposed deep learning approaches (EmbedNet and ChanEstNet), we generate $100{,}000$ dataset samples offline under diverse random channel realizations. Adam with a learning rate $10^{-3}$ is used for a sufficient number of epochs with a batch size of $128$ until validation performance stabilizes. Both networks are trained once at a fixed SNR of 20 dB and then evaluated over the full SNR range without any retraining or fine-tuning. Training at a high SNR allows the models to learn the underlying covariance structure from relatively clean observations, rather than fitting to noise-dominated samples \footnote{  In ablation experiments, we observed that single high SNR training strategy achieves comparable or better BER across the SNR range than models trained only at low SNR or on mixed SNR datasets.}. We conduct $10^5$ Monte Carlo trials for each SNR point to estimate the bit error rates. For each trial, we randomly generate channel coefficients, select pilot or data symbols accordingly, and apply each detection method. We report the resulting bit error rates as a function of SNR.
	
	We also comment on computational complexity and inference time. The optimal LRT with perfect CSI evaluates all $2^N$ hypotheses and has complexity $\mathcal{O}(2^N M K)$ for $M$ antennas and $K$ RF samples, whereas ED has complexity $\mathcal{O}(MK)$. EmbedNet processes all $T$ symbols in a frame through the neural network ($\mathrm{netA}_{\boldsymbol{\theta}}(\cdot)$) and incurs $2^N$ prototype distance computations per data symbol, yielding a complexity of $\mathcal{O}(T \mathrm{netA}_{\boldsymbol{\theta}}(\cdot) + (T-P) 2^N D)$ per frame, where $D$ is the embedding dimension. ChanEstNet runs the neural network ($\mathrm{netB}_{\boldsymbol{\phi}}(\cdot)$) once per frame and then applies LRT, resulting in complexity $\mathcal{O}(\mathrm{netB}_{\boldsymbol{\phi}}(\cdot) + (T-P) 2^N M K)$ per frame. In a representative setting with $N=2$, $M=4$, $K=20$, $P=32$ and $T=160$ per frame, on an NVIDIA RTX 3070 Ti the average inference time per frame is 46.382 ms for the LRT with perfect CSI, 1.279 ms for EmbedNet, 0.282 ms for ChanEstNet in addition to the LRT cost, and 0.281 ms for ED. Although EmbedNet and ChanEstNet also involve $2^N$ terms, their per hypothesis computations are either low-dimensional distance evaluations (EmbedNet) or limited to the final LRT step after channel estimation (ChanEstNet). In contrast, the LRT must evaluate full likelihoods over all $MK$ samples per hypothesis, which explains the much smaller runtime of EmbedNet and ChanEstNet.
	
	\begin{figure}[t]
      \centering
      \begin{tikzpicture}
          \begin{semilogyaxis}[
              width=1\linewidth,
              height=0.9\linewidth,
              xlabel={SNR (dB)},
              xlabel style={font=\small},
              ylabel={BER},
              ylabel style={font=\small},
              y label style={at={(-0.1,0.5)}},
              xmin=0, xmax=20,
              xtick={0, 4, 8, 12, 16, 20},
              ymin=0.0001, ymax=1,
              grid=both,
              grid style={dotted, gray!50},
              minor grid style={dotted, gray!50},
              legend style={at={(0.01,0.01)}, anchor=south west, font=\footnotesize, draw=gray, fill=white},
              legend cell align=left,
              clip=false,
              ]

              \addplot[
                  color=figcolor1,
                  mark=o,
                  line width=1pt,
                  solid,
              ] coordinates {
                  (0, 0.2761) (4, 0.1838) (8, 0.0842)
                  (12, 0.0226) (16, 0.0026) (20, 0.0002)
              };
              \addlegendentry{LRT ($N=2$)}

              \addplot[
                  color=figcolor1,
                  mark=o,
                  mark options={fill=white},
                  line width=1pt,
                  dashed,
              ] coordinates {
                  (0, 0.2824) (4, 0.1903) (8, 0.0916)
                  (12, 0.0252) (16, 0.0033) (20, 0.0003)
              };
              \addlegendentry{LRT ($N=3$)}

              \addplot[
                  color=figcolor2,
                  mark=x,
                  line width=1pt,
                  solid,
              ] coordinates {
                  (0, 0.3931) (4, 0.2909) (8, 0.1668)
                  (12, 0.0657) (16, 0.0157) (20, 0.0028)
              };
              \addlegendentry{EmbedNet ($N=2$)}

              \addplot[
                  color=figcolor2,
                  mark=x,
                  line width=1pt,
                  dashed,
              ] coordinates {
                  (0, 0.4117) (4, 0.3156) (8, 0.1980)
                  (12, 0.0853) (16, 0.0224) (20, 0.0041)
              };
              \addlegendentry{EmbedNet ($N=3$)}

              \addplot[
                  color=figcolor3,
                  mark=triangle,
                  line width=1pt,
                  solid,
              ] coordinates {
                  (0, 0.3468) (4, 0.2400) (8, 0.1246)
                  (12, 0.0442) (16, 0.0116) (20, 0.0028)
              };
              \addlegendentry{ChanEstNet ($N=2$)}

              \addplot[
                  color=figcolor3,
                  mark=triangle,
                  mark options={fill=white},
                  line width=1pt,
                  dashed,
              ] coordinates {
                  (0, 0.3635) (4, 0.2664) (8, 0.1513)
                  (12, 0.0598) (16, 0.0171) (20, 0.0044)
              };
              \addlegendentry{ChanEstNet ($N=3$)}

              \addplot[
                  color=figcolor4,
                  mark=+,
                  line width=1pt,
                  solid,
              ] coordinates {
                  (0, 0.46510) (4, 0.44685) (8, 0.44800)
                  (12, 0.44220) (16, 0.45470) (20, 0.44460)
              };
              \addlegendentry{ED ($N=2$)}

              \addplot[
                  color=figcolor4,
                  mark=+,
                  line width=1pt,
                  dashed,
              ] coordinates {
                  (0, 0.46110) (4, 0.45717) (8, 0.44883)
                  (12, 0.45117) (16, 0.44500) (20, 0.44690)
              };
              \addlegendentry{ED ($N=3$)}

          \end{semilogyaxis}
      \end{tikzpicture}
      \caption{Gaussian ambient source with $\zeta_i = -20$ dB for $N=2$ (solid lines) and $N=3$ (dashed lines).}
      \label{fig:gaussian_combined}
      \vspace{-0.5cm}
  	\end{figure}
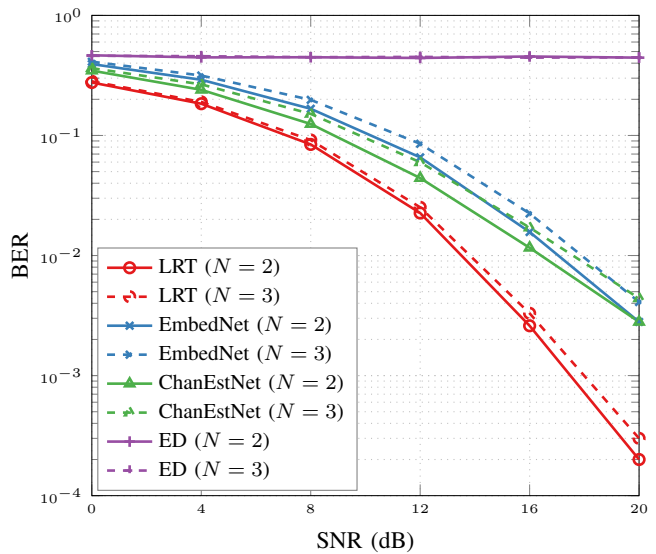	
  	
	Fig. \ref{fig:gaussian_combined} reports the BER performance when the ambient source is Gaussian with $\zeta_i = -20$ dB for $N=2$ and $N=3$. As expected, the LRT with perfect CSI provides the lowest BER across the entire SNR range and serves as a lower bound on achievable BER. ChanEstNet and EmbedNet both closely track this bound and substantially reduce BER. ChanEstNet typically achieves lower BER by refining pilot correlations into explicit channel estimates that are then used in a multi-hypothesis LRT, whereas EmbedNet remains CSI-free with pilot-based prototype adaptation. The ED baseline exhibits limited SNR sensitivity under a strong Gaussian direct path. Increasing the number of tags from two to three shifts all curves upward due to the larger $2^N$ hypothesis space, but the ordering of the curves remains unchanged.
	
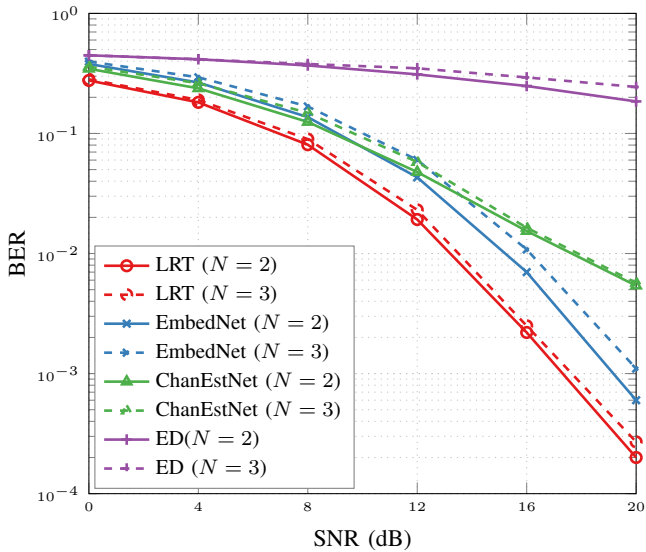
\begin{figure}[t]
      \centering
      \begin{tikzpicture}
          \begin{semilogyaxis}[
              width=1\linewidth,
              height=0.9\linewidth,
              xlabel={SNR (dB)},
              xlabel style={font=\small},
              ylabel={BER},
              ylabel style={font=\small},
              y label style={at={(-0.1,0.5)}},
              xmin=0, xmax=20,
              xtick={0, 4, 8, 12, 16, 20},
              ymin=0.0001, ymax=1,
              grid=both,
              grid style={dotted, gray!50},
              minor grid style={dotted, gray!50},
              legend style={at={(0.01,0.01)}, anchor=south west, font=\footnotesize, draw=gray, fill=white},
              legend cell align=left,
              clip=false,
              ]

              \addplot[
                  color=figcolor1,
                  mark=o,
                  line width=1pt,
                  solid,
              ] coordinates {
                  (0, 0.2763) (4, 0.1815) (8, 0.0806)
                  (12, 0.0192) (16, 0.0022) (20, 0.0002)
              };
              \addlegendentry{LRT ($N=2$)}

              \addplot[
                  color=figcolor1,
                  mark=o,
                  mark options={fill=white},
                  line width=1pt,
                  dashed,
              ] coordinates {
                  (0, 0.2825) (4, 0.1901) (8, 0.0897)
                  (12, 0.0231) (16, 0.0025) (20, 0.00027)
              };
              \addlegendentry{LRT ($N=3$)}

              \addplot[
                  color=figcolor2,
                  mark=x,
                  line width=1pt,
                  solid,
              ] coordinates {
                  (0, 0.3790) (4, 0.2657) (8, 0.1368)
                  (12, 0.0431) (16, 0.0070) (20, 0.0006)
              };
              \addlegendentry{EmbedNet ($N=2$)}

              \addplot[
                  color=figcolor2,
                  mark=x,
                  line width=1pt,
                  dashed,
              ] coordinates {
                  (0, 0.3973) (4, 0.2935) (8, 0.1679)
                  (12, 0.0602) (16, 0.0108) (20, 0.0011)
              };
              \addlegendentry{EmbedNet ($N=3$)}

              \addplot[
                  color=figcolor3,
                  mark=triangle,
                  line width=1pt,
                  solid,
              ] coordinates {
                  (0, 0.3458) (4, 0.2376) (8, 0.1247)
                  (12, 0.0477) (16, 0.0154) (20, 0.0054)
              };
              \addlegendentry{ChanEstNet ($N=2$)}

              \addplot[
                  color=figcolor3,
                  mark=triangle,
                  mark options={fill=white},
                  line width=1pt,
                  dashed,
              ] coordinates {
                  (0, 0.3598) (4, 0.2620) (8, 0.1486)
                  (12, 0.0578) (16, 0.0164) (20, 0.0057)
              };
              \addlegendentry{ChanEstNet ($N=3$)}

              \addplot[
                  color=figcolor4,
                  mark=+,
                  line width=1pt,
                  solid,
              ] coordinates {
                  (0, 0.44755) (4, 0.41510) (8, 0.36855)
                  (12, 0.31055) (16, 0.24840) (20, 0.18490)
              };
              \addlegendentry{ED($N=2$)}

              \addplot[
                  color=figcolor4,
                  mark=+,
                  line width=1pt,
                  dashed,
              ] coordinates {
                  (0, 0.44813) (4, 0.41457) (8, 0.37893)
                  (12, 0.34890) (16, 0.29293) (20, 0.24390)
              };
              \addlegendentry{ED ($N=3$)}

          \end{semilogyaxis}
      \end{tikzpicture}
      \caption{QPSK ambient source with $\zeta_i = -20$ dB for $N=2$ (solid lines) and $N=3$ (dashed lines).}
      \label{fig:qpsk_combined}
      \vspace{-0.5cm}
  	\end{figure}

	Fig. \ref{fig:qpsk_combined} depicts the BER performance under a QPSK ambient source\footnote{We have also validated our proposed methods under a 16-QAM ambient source with similar performance trends.} for $N=2$ and $N=3$. Similarly, the LRT with perfect CSI provides the lowest BER and serves as a reference. Compared to the Gaussian case, all methods benefit from the discrete QPSK structure, yielding noticeably lower BER across the SNR range. Unlike the Gaussian case, EmbedNet typically achieves slightly lower BER than ChanEstNet under QPSK for high SNR values, likely because the constant modulus and discrete phase properties of QPSK yield more consistent pilot-based prototypes, whereas residual channel estimation errors can limit ChanEstNet in the subsequent multi-hypothesis LRT. The ED baseline improves with SNR but remains substantially worse than the proposed deep learning schemes. Increasing the number of tags from two to three again shifts all curves upward but the ordering of the curves remains unchanged.
	
\begin{figure}[t]
		\centering
		\begin{tikzpicture}
			\begin{semilogyaxis}[
				width=1\linewidth,
				height=0.9\linewidth,
				xlabel={$\zeta_i$ (dB)},
				xlabel style={font=\small},
				ylabel={BER},
				ylabel style={font=\small},
				y label style={at={(-0.1,0.5)}},
				xmin=-20, xmax=0,
				xtick={-20,-16,-12,-8,-4,0},
				ymin=0.0001, ymax=1,
				grid=both,
				grid style={dotted, gray!50},
				minor grid style={dotted, gray!50},
				legend style={at={(0.01,0.01)}, anchor=south west, font=\small, draw=gray, fill=white},
				legend cell align=left,
				clip=false,
				]
				\addplot[
				color=figcolor1,
				mark=o,
				line width=1pt,
				] coordinates {
					(-20, 0.2779)
					(-16, 0.1822)
					(-12, 0.0836)
					(-8,  0.0200)
					(-4,  0.0023)
					(0,   0.00010)
				};
				\addlegendentry{LRT}
				
				\addplot[
				color=figcolor2,
				mark=x,
				line width=1pt,
				] coordinates {
					(-20, 0.3782)
					(-16, 0.2819)
					(-12, 0.1613)
					(-8, 0.0601)
					(-4, 0.0112)
					(0, 0.0012)
				};
				\addlegendentry{EmbedNet}
				
				\addplot[
				color=figcolor3,
				mark=triangle,
				line width=1pt,
				] coordinates {
					(-20, 0.3458)
					(-16, 0.2368)
					(-12, 0.1139)
					(-8, 0.0303)
					(-4, 0.0040)
					(0, 0.0002)
				};
				\addlegendentry{ChanEstNet}
			\end{semilogyaxis}
		\end{tikzpicture}
		\caption{QPSK ambient source, $\text{SNR} = 0$ dB, $K=20$,  and $N=2$.}
		\label{fig9}
		\vspace{-0.5cm}
	\end{figure}
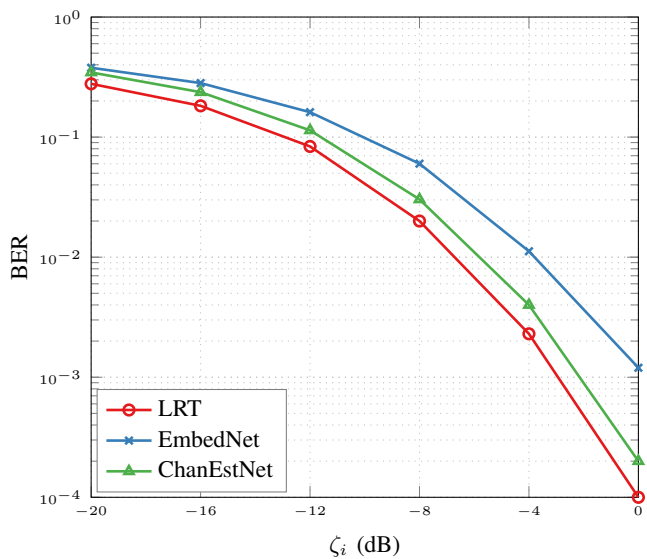

	We also examine how the relative coefficient between the direct and backscatter paths $\zeta_i$ affects BER when the ambient source follows QPSK modulation and $N=2$ tags are present. The BER results are illustrated in Fig. \ref{fig9}. As $\zeta_i$ increases, i.e., the backscatter component becomes stronger relative to the direct path, the BER decreases for all schemes. The LRT with perfect CSI consistently yields the lowest BER. ChanEstNet approaches this benchmark as $\zeta_i$ increases, since stronger reflections improve the reliability of channel refinement using pilot symbols. EmbedNet also benefits from larger $\zeta_i$ but maintains a larger gap to the reference. Overall, strengthening the backscatter path improves detectability, while a residual gap to the ideal LRT remains because the pilots provide only limited and noisy information for estimating the channels or prototypes used for detection.
	
\begin{figure}[t]
		\centering
		\begin{tikzpicture}
			\begin{semilogyaxis}[
				width=1\linewidth,
				height=0.9\linewidth,
				xlabel={$K$ (number of RF source samples)},
				xlabel style={font=\small},
				ylabel={BER},
				ylabel style={font=\small},
				y label style={at={(-0.1,0.5)}},
				xmin=1, xmax=100,
				xtick={1,20,40,60,80,100},
				ymin=0.001, ymax=1,
				grid=both,
				grid style={dotted, gray!50},
				minor grid style={dotted, gray!50},
				legend style={at={(0.01,0.01)}, anchor=south west, font=\small, draw=gray, fill=white},
				legend cell align=left,
				clip=false,
				]
				\addplot[
				color=figcolor1,
				mark=o,
				line width=1pt,
				] coordinates {
					(1, 0.3327)
					(20, 0.0451)
					(40, 0.0129)
					(60, 0.0047)
					(80, 0.0021)
					(100, 0.0012)
				};
				
				\addlegendentry{LRT}
				
				\addplot[
				color=figcolor2,
				mark=x,
				line width=1pt,
				] coordinates {
					(1, 0.4124)
					(20, 0.0875)
					(40, 0.0366)
					(60, 0.0178)
					(80, 0.0097)
					(100, 0.0057)
				};
				\addlegendentry{EmbedNet}
				
				\addplot[
				color=figcolor3,
				mark=triangle,
				line width=1pt,
				] coordinates {
					(1, 0.4033)
					(20, 0.0825)
					(40, 0.0339)
					(60, 0.0161)
					(80, 0.0092)
					(100, 0.0052)
				};
				\addlegendentry{ChanEstNet}
				
			\end{semilogyaxis}
		\end{tikzpicture}
		\caption{QPSK ambient source, $\zeta_i = -20$ dB, $\text{SNR} = 10$ dB, and $N=2$.}
		\label{fig10}
		\vspace{-0.5cm}
	\end{figure}
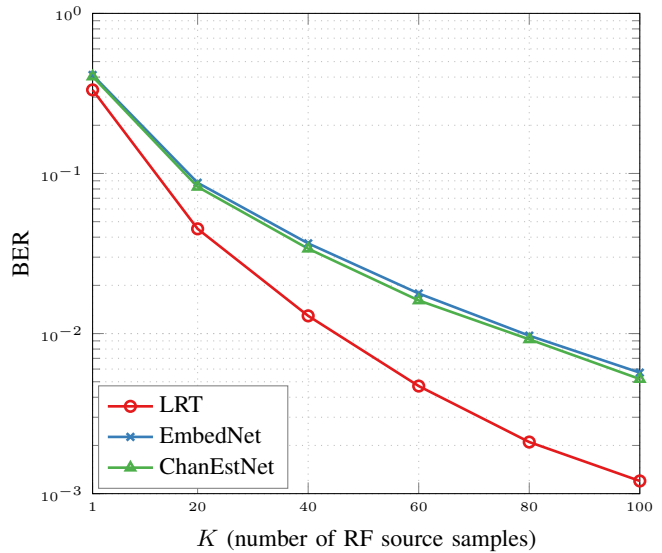

	We further investigate the impact of the number of RF source samples per tag symbol by varying $K$ at $10$ dB SNR for $N=2$ tags under a QPSK ambient source, as shown in Fig. \ref{fig10}. As $K$ increases, each tag symbol is sampled more extensively and the BER of all three detectors decreases. The LRT with perfect CSI benefits the most and approaches very low error rates even for moderate $K$. Both EmbedNet and ChanEstNet exhibit monotonic improvement as $K$ grows, with ChanEstNet typically achieving slightly lower BER than EmbedNet for the considered values of $K$ due to more accurate channel estimation when more samples are available. Overall, increasing $K$ enhances multi-tag detection reliability for all methods, but at the expense of higher per symbol sampling and processing overhead, illustrating a trade-off between complexity and performance in AmBC systems.
	
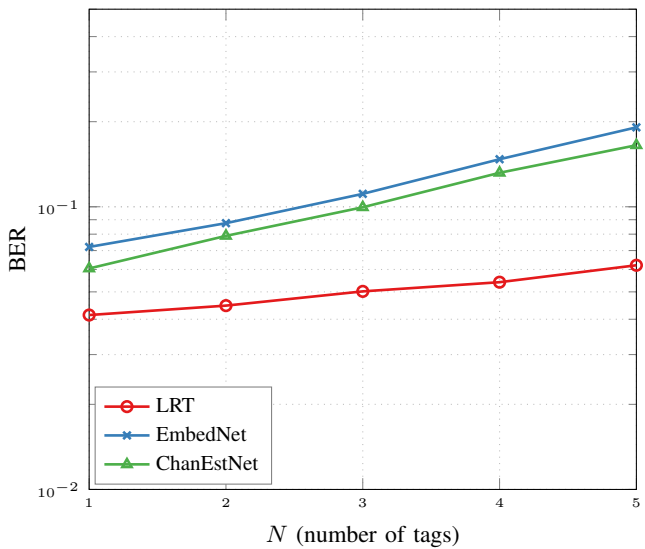
\begin{figure}[t]
      \centering
      \begin{tikzpicture}
          \begin{semilogyaxis}[
              width=1\linewidth,
              height=0.9\linewidth,
              xlabel={$N$ (number of tags)},
              xlabel style={font=\small},
              ylabel={BER},
              ylabel style={font=\small},
              y label style={at={(-0.1,0.5)}},
              xmin=1, xmax=5,
              xtick={1,2,3,4,5},
              ymin=0.01, ymax=0.5,
              grid=both,
              grid style={dotted, gray!50},
              minor grid style={dotted, gray!50},
              legend style={at={(0.01,0.01)}, anchor=south west, font=\footnotesize, draw=gray, fill=white},
              legend cell align=left,
              clip=false,
          ]

          \addplot[
              color=figcolor1,
              mark=o,
              line width=1pt,
              solid,
          ] coordinates {
              (1, 0.0414)
              (2, 0.0447)
              (3, 0.0502)
              (4, 0.0541)
              (5, 0.0621)
          };
          \addlegendentry{LRT}

          \addplot[
              color=figcolor2,
              mark=x,
              line width=1pt,
              solid,
          ] coordinates {
              (1, 0.0721)
              (2, 0.0875)
              (3, 0.1111)
              (4, 0.1473)
              (5, 0.1910)
          };
          \addlegendentry{EmbedNet}

          \addplot[
              color=figcolor3,
              mark=triangle,
              line width=1pt,
              solid,
          ] coordinates {
              (1, 0.0605)
              (2, 0.0789)
              (3, 0.0996)
              (4, 0.1319)
              (5, 0.1651)
          };
          \addlegendentry{ChanEstNet}

          \end{semilogyaxis}
      \end{tikzpicture}
      \caption{QPSK ambient source, $\text{SNR}=10$ dB and $\zeta_i=-20$ dB.}
      \label{fig:scalability_N}
  	\end{figure}

	Fig. \ref{fig:scalability_N} illustrates the BER results for different number of tags $N$ under a QPSK ambient source for $\text{SNR}=10$ dB and $\zeta_i=-20$ dB. As $N$ increases from $1$ to $5$, the BER increases monotonically for all detectors. The LRT with perfect CSI remains the best across all $N$, while ChanEstNet consistently lies between LRT and EmbedNet in this setting. Although most of our earlier results focus on $N=2$ and $N=3$, both proposed deep learning architectures remain effective up to at least $N=5$. The same designs can be applied to larger $N$ in theory, but the exponential growth of the $2^N$ hypothesis space and the need to allocate sufficient pilots mean that pilot overhead and detection complexity will ultimately limit the achievable number of tags in practice.

\begin{figure}[t]
      \centering
      \begin{tikzpicture}
          \begin{semilogyaxis}[
              width=1\linewidth,
              height=0.9\linewidth,
              xlabel={$P$ (number of pilot symbols)},
              xlabel style={font=\small},
              ylabel={BER},
              ylabel style={font=\small},
              y label style={at={(-0.1,0.5)}},
              xmin=8, xmax=64,
              xtick={8,16,32,64},
              ymin=0.01, ymax=0.5,
              grid=both,
              grid style={dotted, gray!50},
              minor grid style={dotted, gray!50},
              legend style={at={(0.01,0.01)}, anchor=south west,
                            font=\footnotesize, draw=gray, fill=white,  legend columns = 2},
              legend cell align=left,
              clip=false,
          ]

          \addplot[
              color=figcolor1,
              mark=o,
              line width=1pt,
              solid,
          ] coordinates {
              (8,  0.0443)
              (16, 0.0443)
              (32, 0.0443)
              (64, 0.0443)
          };
          \addlegendentry{LRT ($N=2$)}

          \addplot[
              color=figcolor1,
              mark=o,
              mark options={fill=white},
              line width=1pt,
              dashed,
          ] coordinates {
              (8,  0.0495)
              (16, 0.0495)
              (32, 0.0495)
              (64, 0.0495)
          };
          \addlegendentry{LRT ($N=3$)}

          \addplot[
              color=figcolor2,
              mark=x,
              line width=1pt,
              solid,
          ] coordinates {
              (8,  0.1315)
              (16, 0.1041)
              (32, 0.0869)
              (64, 0.0803)
          };
          \addlegendentry{EmbedNet ($N=2$)}

          \addplot[
              color=figcolor2,
              mark=x,
              line width=1pt,
              dashed,
          ] coordinates {
              (8,  0.1870)
              (16, 0.1394)
              (32, 0.1102)
              (64, 0.0963)
          };
          \addlegendentry{EmbedNet ($N=3$)}

          \addplot[
              color=figcolor3,
              mark=triangle,
              line width=1pt,
              solid,
          ] coordinates {
              (8,  0.1243)
              (16, 0.0925)
              (32, 0.0814)
              (64, 0.0695)
          };
          \addlegendentry{ChanEstNet ($N=2$)}

          \addplot[
              color=figcolor3,
              mark=triangle,
              mark options={fill=white},
              line width=1pt,
              dashed,
          ] coordinates {
              (8,  0.1670)
              (16, 0.1281)
              (32, 0.1014)
              (64, 0.0896)
          };
          \addlegendentry{ChanEstNet ($N=3$)}

          \end{semilogyaxis}
      \end{tikzpicture}
      \caption{QPSK ambient source, $\text{SNR}=10$ dB and $\zeta_i=-20$ dB for $N=2$ and $N=3$ tags.}
      \label{fig:pilot_scaling}
      \vspace{-0.5cm}
  	\end{figure}
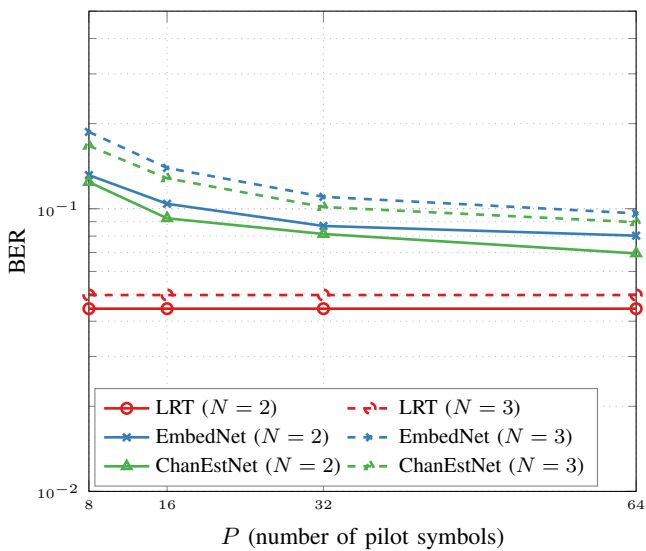

	 Fig. \ref{fig:pilot_scaling} examines the impact of the number of pilot symbols $P$ on BER for a QPSK ambient source at $\text{SNR}=10$ dB and $\zeta_i=-20$ dB with $N=2$ and $N=3$ tags. The LRT benchmark assumes perfect CSI and therefore remains essentially independent of $P$, appearing flat for both values of $N$. In contrast, both proposed detectors improve as $P$ increases because additional pilots reduce the noise in the pilot information used for detection. Specifically, EmbedNet benefits through more reliable prototypes in the embedding space, whereas ChanEstNet benefits through more accurate effective channel estimates that feed the multi-hypothesis LRT. Increasing $P$ from $8$ to $32$ yields a substantial BER reduction, while further increasing to $P=64$ provides diminishing gains. These results highlight the trade-off between pilot overhead and detection performance, since larger $P$ reduces the number of symbols available for data.
  	
	In order to place EmbedNet in context, we report a normalized throughput per tag that accounts for pilot overhead and compare it with the non-coherent parallel detection scheme in \cite{Yang2023Non-Coherent} and with the LRT benchmark with perfect CSI for $N=2$ and $N=3$ tags. These approaches adopt different signaling overheads, with pilot symbols in our framework and FM0/Miller line coding without explicit pilots in \cite{Yang2023Non-Coherent}. They are evaluated under their respective system models, the resulting throughput values should be interpreted as a qualitative reference at the system level. Table \ref{tab:throughput_snr} summarizes the corresponding normalized throughput across SNR.

	In order to compare methods in terms of overall efficiency, we convert the BER values into a normalized throughput per tag. For a frame with $T$ symbols and $P$ pilots, the fraction of symbols available for data is $\eta_{\text{data}} = 1 - P/T$. We define $b_{\mathrm{pc}}$ as number of bits per tag symbol, the normalized throughput at a given SNR is defined as $\mathcal{T}_{\text{tag}}(\mathrm{SNR}) = b_{\mathrm{pc}}\,\eta_{\text{data}}\,\bigl(1 - \mathrm{BER}(\mathrm{SNR})\bigr)$. In our setting, we use $T=160$ and $P=32$, so $\eta_{\text{data}} = 0.8$, and on-off keying yields $b_{\mathrm{pc}}^{\text{EmbedNet}} = 1$, leading to $\mathcal{T}_{\text{tag}}^{\text{EmbedNet}}(\mathrm{SNR}) = 0.8\bigl(1 - \mathrm{BER}_{\text{EmbedNet}}(\mathrm{SNR})\bigr)$. For the LRT benchmark, we apply the same payload fraction $\eta_{\text{data}} = 0.8$ to ensure consistent overhead accounting, i.e., $\mathcal{T}_{\text{tag}}^{\text{LRT}}(\mathrm{SNR}) = 0.8\bigl(1 - \mathrm{BER}_{\text{LRT}}(\mathrm{SNR})\bigr)$. For the detector in \cite{Yang2023Non-Coherent}, we assume no explicit pilot symbols (thus $\eta_{\text{data}}^{\text{Yang}} = 1$), and account for the FM0/Miller line coding by adopting an effective $b_{\mathrm{pc}}^{\text{Yang}} = 0.5$, since one information bit is conveyed by two symbol intervals. This gives $\mathcal{T}_{\text{tag}}^{\text{Yang}}(\mathrm{SNR}) = 0.5\bigl(1 - \mathrm{BER}_{\text{Yang}}(\mathrm{SNR})\bigr)$. The throughput entries in Table \ref{tab:throughput_snr} are obtained by applying these expressions to the BER values from our simulations and from \cite{Yang2023Non-Coherent} for each SNR and each method.
		
  \begin{table}[]
      \footnotesize
      \centering
      \caption{Normalized per tag throughput comparison of LRT benchmark, EmbedNet,  and Yang \cite{Yang2023Non-Coherent}}
      \label{tab:throughput_snr}
      \renewcommand{\arraystretch}{1.5}
      \setlength{\tabcolsep}{5pt}
      
      \begin{tabular}{|c|cc|cc|cc|}
          \hline
	          \textbf{SNR (dB)} & \multicolumn{2}{c|}{\textbf{LRT}} & \multicolumn{2}{c|}{\textbf{EmbedNet}} & \multicolumn{2}{c|}{\textbf{Yang \cite{Yang2023Non-Coherent}}} \\
           & \textbf{$N=2$} & \textbf{$N=3$} & \textbf{$N=2$} & \textbf{$N=3$} & \textbf{$N=2$} & \textbf{$N=3$} \\
          \hline
          $-5$ & 0.573 & 0.572 & 0.467 & 0.457 & 0.385 & 0.325 \\ \hline
           0   & 0.670 & 0.665 & 0.592 & 0.568 & 0.453 & 0.409 \\ \hline
           5   & 0.761 & 0.757 & 0.726 & 0.704 & 0.484 & 0.460 \\ \hline
          10   & 0.796 & 0.796 & 0.790 & 0.785 & 0.495 & 0.481 \\ \hline
          15   & 0.800 & 0.800 & 0.799 & 0.799 & 0.498 & 0.493 \\ \hline
          20   & 0.800 & 0.800 & 0.800 & 0.800 & 0.499 & 0.498 \\ \hline
          25   & 0.800 & 0.800 & 0.800 & 0.800 & 0.500 & 0.499 \\ \hline
      \end{tabular}
      \vspace{-0.5cm}
  \end{table}
  
    For our schemes and the LRT benchmark, the maximum achievable throughput is limited by the pilot overhead to $0.8$, whereas for the detector in \cite{Yang2023Non-Coherent} the ceiling is $0.5$ due to the FM0/Miller line coding overhead. As SNR increases, both the LRT and the proposed method rapidly approach the pilot limited ceiling of $0.8$, while the throughput of \cite{Yang2023Non-Coherent} saturates near $0.5$ despite further SNR improvements. Consequently, in the moderate and high SNR regime, the proposed deep learning framework provides a substantially higher effective throughput per tag and remains quantitatively closer to the LRT benchmark.

	\section{Conclusions}
	In this paper, we have presented a comprehensive study on deep learning based signal detection for multi-tag ambient backscatter communication systems. Our work addresses the inherent challenges posed by low-power backscatter signals and multi-hypothesis detection in scenarios with multiple passive tags. First, we established theoretical bounds on detection performance using a multi-hypothesis likelihood ratio test (LRT), the pairwise error probability, and energy detection. We then introduced two deep learning driven approaches to relax the requirement of explicit channel state information. The first approach employs a prototype based classifier wherein the embedding network constructs class prototypes for each tag state combination and adaptively recovers the tag bits from pilot symbols, reducing complexity and eliminating the need for a channel estimation. The second approach uses a dedicated neural network to obtain channel estimates from limited pilot observations before performing an LRT on the data symbols. By incorporating classical detection theory, this hybrid scheme balances the interpretability of model based detectors and the versatility of deep learning. Extensive simulations under varying system parameters validate the accuracy and robustness of both proposed methods. 
	
	Future research directions include extending these techniques to large scale tag deployments, investigating their efficacy under fast fading or mobility, and exploring advanced neural architectures or distributed training strategies. Real-world implementation can also reveal practical considerations such as hardware imperfections, pilot overhead constraints, and on-chip resource limitations.
	
	\bibliographystyle{IEEEtran}
	\bibliography{ref}

@article{Huynh2017Ambient,
	title={Ambient Backscatter Communications: A Contemporary Survey},
	author={Nguyen Van Huynh and D. Hoang and Xiao Lu and D. Niyato and Ping Wang and Dong In Kim},
	journal={IEEE Commun. Surveys  Tuts.},
	year={2018},
	volume={20},
	pages={2889-2922},
}

@article{Lu2018Ambient,
	title={Ambient Backscatter Assisted Wireless Powered Communications},
	author={Xiao Lu and D. Niyato and Hai Jiang and Dong In Kim and Yong Xiao and Zhu Han},
	journal={IEEE Wireless Commun.},
	year={2018},
	volume={25},
	number  = {2},
	pages={170-177},
	doi={10.1109/MWC.2017.1600398}
}

@article{niu2019overview,
	title={An overview on backscatter communications},
	author={Niu, Jin-Ping and Li, Geoffrey Ye},
	journal={J. Commun. Inf. Netw.},
	volume={4},
	number={2},
	pages={1--14},
	year={2019},
	publisher={PTP}
}

@article{atzori2010internet,
	title={The internet of things: A survey},
	author={Atzori, Luigi and Iera, Antonio and Morabito, Giacomo},
	journal={Comput. Netw.},
	volume={54},
	number={15},
	pages={2787--2805},
	year={2010},
	publisher={Elsevier}
}

@inproceedings{rose2015internet,
	title={The internet of things: An overview},
	author={Rose, Karen and Eldridge, Scott and Chapin, Lyman},
	booktitle={Proc. Internet Soc. (ISOC)},
	pages={1--53},
	year={2015},

}

@article{li2015internet,
	title={The internet of things:{A} survey},
	author={Li, Shancang and , Li Da and Zhao, Shanshan},
	journal={Inf. Syst. Front},
	volume={17},
	number={2},
	pages={243--259},
	year={2015},
}

@article{duan2020ambient,
	title={Ambient backscatter communications for future ultra-low-power machine type communications: Challenges, solutions, opportunities, and future research trends},
	author={Duan, Ruifeng and Wang, Xiyu and Yigitler, Huseyin and Sheikh, Muhammad Usman and Jantti, Riku and Han, Zhu},
	journal={IEEE Commun. Mag.},
	volume={58},
	number={2},
	pages={42--47},
	year={2020},
	publisher={IEEE}
}

@inproceedings{liu2013ambient,
	title={Ambient backscatter: Wireless communication out of thin air},
	author={Liu, Vincent and Parks, Aaron and Talla, Vamsi and Gollakota, Shyamnath and Wetherall, David and Smith, Joshua R},
	booktitle={Proc. ACM SIGCOMM},
	pages={39--50},
	address = {Hong Kong},
	year={2013},
	month = {Aug.},
}

@article{chowdhury20206g,
	title={{6G} wireless communication systems: Applications, requirements, technologies, challenges, and research directions},
	author={Chowdhury, Mostafa Zaman and Shahjalal, Md and Ahmed, Shakil and Jang, Yeong Min},
	journal={IEEE Open  J. Commun. Soc.},
	volume={1},
	pages={957--975},
	year={2020},
	publisher={IEEE}
}

@article{akyildiz20206g,
	title={{6G} and beyond: The future of wireless communications systems},
	author={Akyildiz, Ian F and Kak, Ahan and Nie, Shuai},
	journal={IEEE Access},
	volume={8},
	pages={133995--134030},
	year={2020},
	publisher={IEEE}
}

@article{Wang2016Ambient,
	title={Ambient Backscatter Communication Systems: Detection and Performance Analysis},
	author={Gongpu Wang and F. Gao and Rongfei Fan and C. Tellambura},
	journal={IEEE Trans. Commun.},
	year={2016},
	number={11},
	volume={64},
	pages={4836-4846},
}

@article{Qian2017Noncoherent,
	title={Noncoherent Detections for Ambient Backscatter System},
	author={Jing Qian and F. Gao and Gongpu Wang and Shi Jin and Hongbo Zhu {}},
	journal={IEEE Trans. Wireless Commun.},
	year={2017},
	volume={16},
	number={3},
	pages={1412-1422},
}

@article{Yang2018Cooperative,
	title={Cooperative Ambient Backscatter Communications for Green Internet-of-Things},
	author={Gang Yang and Qianqian Zhang and Ying-Chang Liang},
	journal={IEEE Internet Things J.},
	year={2018},
	volume={5},
	number={2},
	pages={1116-1130},
}

@article{Darsena2019Noncoherent,
	title={Noncoherent Detection for Ambient Backscatter Communications Over {OFDM} Signals},
	author={D. Darsena},
	journal={IEEE Access},
	year={2019},
	volume={7},
	pages={159415-159425},
}

@article{Liu2017Coding,
	title={Coding and Detection Schemes for Ambient Backscatter Communication Systems},
	author={Yang Liu and Gongpu Wang and Zhongzhao Dou and Z. Zhong},
	journal={IEEE Access},
	year={2017},
	volume={5},
	pages={4947-4953},
}

@article{Qian2016Semi-Coherent,
	title={Semi-Coherent Detection and Performance Analysis for Ambient Backscatter System},
	author={Jing Qian and F. Gao and Gongpu Wang and Shi Jin and Hongbo Zhu},
	journal={IEEE Trans. Commun.},
	year={2017},
	volume={65},
	number={12},
	pages={5266-5279},
}

@article{Elmossallamy2019Noncoherent,
	title={Noncoherent Backscatter Communications Over Ambient {OFDM} Signals},
	author={Mohamed A. Elmossallamy and M. Pan and Riku Jäntti and Karim G. Seddik and Geoffrey Y. Li and Zhu Han},
	journal={IEEE Trans. Commun.},
	year={2019},
	volume={67},
	number={5},
	pages={3597-3611},
}

@article{Yang2023Non-Coherent,
	title={Non-Coherent Parallel Detection of Ambient Backscatter Communications With Multiple Tags},
	author={Gang Yang and Zhiyi Luo and Ning Jin and Ying-Chang Liang and Yongjun Xu and Gongpu Wang},
	journal={IEEE Trans. Veh. Technol.},
	year={2023},
	volume={72},
	number={4},
	pages={5344-5349},
	doi={10.1109/TVT.2022.3222005}
}

@article{Zhang2019Constellation,
	title={Constellation Learning-Based Signal Detection for Ambient Backscatter Communication Systems},
	author={Qianqian Zhang and Huayan Guo and Ying-Chang Liang and Xiaojun Yuan},
	journal={IEEE J. Sel. Areas Commun.},
	year={2019},
	volume={37},
	number={2},
	pages={452-463},
	doi={10.1109/JSAC.2018.2872382}
}

@inproceedings{Hu2019Machine,
	title={Machine Learning Based Signal Detection for Ambient Backscatter Communications},
	author={Yunkai Hu and Peng Wang and Zihuai Lin and Ming Ding and Ying-Chang Liang},
	booktitle={Proc. IEEE Int. Conf. Commun. (ICC)},
	address = {Shanghai, China},
	month  = {Jul.},
	year={2019},
}

@inproceedings{Zhang2017Signal,
	title={Signal Detection for Ambient Backscatter Communications Using Unsupervised Learning},
	author={Qianqian Zhang and Ying-Chang Liang},
	booktitle={Proc. IEEE Globecom Workshops (GC Wkshps)},
	month  = {Dec.},
	year={2017},
}

@inproceedings{wang2019machine,
	title={Machine learning-assisted detection for {BPSK}-modulated ambient backscatter communication systems},
	author={Wang, Xiyu and Duan, Ruifeng and Yigitler, Huseyin and Menta, Estifanos and Jantti, Riku},
	booktitle={Proc. IEEE Global Commun. Conf. (GLOBECOM)},
	address = {Waikoloa, HI, USA},
	month  = {Dec.},
	year={2019},
}

@article{Guo2019Exploiting,
	title={Exploiting Multiple Antennas for Cognitive Ambient Backscatter Communication},
	author={Huayan Guo and Qianqian Zhang and Sa Xiao and Ying-Chang Liang},
	journal={IEEE Internet Things J.},
	year={2019},
	month={Feb.},
	volume={6},
	number={1},
	pages={765-775},
}

@article{liu2020deep,
	title={Deep residual learning-assisted channel estimation in ambient backscatter communications},
	author={Liu, Xuemeng and Liu, Chang and Li, Yonghui and Vucetic, Branka and Ng, Derrick Wing Kwan},
	journal={IEEE Wireless Commun. Lett.},
	volume={10},
	number={2},
	pages={339--343},
	month={Feb.},
	year={2021},
	publisher={IEEE}
}

@article{liu2021deep,
	title={Deep transfer learning for signal detection in ambient backscatter communications},
	author={Liu, Chang and Wei, Zhiqiang and Ng, Derrick Wing Kwan and Yuan, Jinhong and Liang, Ying-Chang},
	journal={IEEE Trans. Wireless Commun.},
	volume={20},
	number={3},
	pages={1624--1638},
	month={Mar.},
	year={2021},
	publisher={IEEE}
}

@article{Jia2021IRS-Assisted,
	title={{IRS}-Assisted Ambient Backscatter Communications Utilizing Deep Reinforcement Learning},
	author={Xiaolun Jia and Xiangyun Zhou},
	journal={IEEE Wireless Commun. Lett.},
	year={2021},
	volume={10},
	number={11},
	month={Nov.},
	pages={2374-2378},
	doi={10.1109/lwc.2021.3100901}
}

@inproceedings{jameel2020low,
	title={Low latency ambient backscatter communications with deep {Q}-learning for beyond {5G} applications},
	author={Jameel, Furqan and Jamshed, Muhammad Ali and Chang, Zheng and J{\"a}ntti, Riku and Pervaiz, Haris},
	booktitle={Proc. IEEE 91st Veh. Technol. Conf. (VTCSpring)},
	month={May.},
	address = {Antwerp, Belgium},
	year={2020},
}

@article{toro2021backscatter,
	title={Backscatter wireless communications and sensing in green {Internet of Things}},
	author={Toro, Usman Saleh and Wu, Kaishun and Leung, Victor CM},
	journal={IEEE Trans. Green Commun. Netw.},
	volume={6},
	number={1},
	pages={37--55},
	month={Mar.},
	year={2022},
	publisher={IEEE}
}

@article{zargari2023signal,
	title={Signal detection in ambient backscatter systems: Fundamentals, methods, and trends},
	author={Zargari, Shayan and Hakimi, Azar and Rezaei, Fatemeh and Tellambura, Chintha and Maaref, Amine},
	journal={IEEE Access},
	volume={11},
	pages={140287–-140324},
	year={2023},
	publisher={IEEE}
}

@inproceedings{akyildiz2022ml,
	title={{ML}-aided collision recovery for {UHF-RFID} systems},
	author={Aky{\i}ld{\i}z, Talha and Ku, Raymond and Harder, Nicholas and Ebrahimi, Najme and Mahdavifar, Hessam},
	booktitle={Proc. IEEE Int. Conf. RFID (RFID)},
	address = {Las Vegas, NV, USA}, 
	month={May.},
	year={2022},
}

@article{jiang2023backscatter,
	title={Backscatter communication meets practical battery-free Internet of Things: A survey and outlook},
	author={Jiang, Tao and Zhang, Yu and Ma, Wenyuan and Peng, Miaoran and Peng, Yuxiang and Feng, Mingjie and Liu, Guanghua},
	journal={IEEE Commun. Surveys Tut.},
	volume={25},
	number={3},
	pages={2021--2051},
	year={2023},
	publisher={IEEE}
}

@article{xu2023state,
	title={The state of {AI}-empowered backscatter communications: A comprehensive survey},
	author={Xu, Fang and Hussain, Touseef and Ahmed, Manzoor and Ali, Khurshed and Mirza, Muhammad Ayzed and Khan, Wali Ullah and Ihsan, Asim and Han, Zhu},
	journal={Internet Things J.},
	volume={10},
	number={24},
	pages={21763–-21786},
	month={Dec.},
	year={2023},
	publisher={IEEE}
}

@article{xu2018practical,
	title={Practical backscatter communication systems for battery-free Internet of Things: A tutorial and survey of recent research},
	author={Xu, Chenren and Yang, Lei and Zhang, Pengyu},
	journal={IEEE Signal Process. Mag.},
	volume={35},
	number={5},
	pages={16--27},
	month={Sep.},
	year={2018},
	publisher={IEEE}
}

@article{zargari2024deep,
	title={Deep conditional generative adversarial networks for efficient channel estimation in {AmBC} systems},
	author={Zargari, Shayan and Tellambura, CHINTHA and Maaref, Amine and Li, Geoffrey Ye},
	journal={IEEE Trans.  Mach. Learn. Commun. Netw.},
	volume={2},
	pages={805--822},
	year={2024},
	publisher={IEEE}
}

@article{zargari2024enhancing,
	title={Enhancing {AmBC} systems with deep learning for joint channel estimation and signal detection},
	author={Zargari, Shayan and Hakimi, Azar and Tellambura, Chintha and Maaref, Amine},
	journal={IEEE Trans.  Commun.},
	volume={72},
	number={3},
	pages={1716--1731},
	year={2024},
	publisher={IEEE}
}

@inproceedings{snell2017prototypical,
	author={Snell, Jake and Swersky, Kevin and Zemel, Richard},
	title={Prototypical networks for few-shot learning},
	booktitle = {Proc. 31st Conf. Neural Inf. Process. Syst.,},
	year={2017},
	pages={4077-4087},
}

@article{kingma2014adam,
	title={Adam: A method for stochastic optimization},
	author={Kingma, Diederik P and Ba, Jimmy},
	journal={arXiv preprint arXiv:1412.6980},
	year={2014}
}

@book{jacobs1965principles,
	title={Principles of communication engineering.},
	author={Jacobs, Irwin Mark and Wozencraft, JM},
	year={1965},
	publisher  = {Wiley},
	address = {New York},
}

@article{yang2017modulation,
  title={Modulation in the air: Backscatter communication over ambient {OFDM} carrier},
  author={Yang, Gang and Liang, Ying-Chang and Zhang, Rui and Pei, Yiyang},
  journal={IEEE Trans.  Commun.},
  volume={66},
  number={3},
  pages={1219--1233},
  year={2018},
  publisher={IEEE}
}

@inproceedings{parks2014turbocharging,
  title={Turbocharging ambient backscatter communication},
  author={Parks, Aaron N and Liu, Angli and Gollakota, Shyamnath and Smith, Joshua R},
  booktitle={Proc. ACM SIGCOMM},
  pages={619--630},
  year={2014},
  address = {Chicago, IL, USA}, 
  month={Aug.},
  year={2014},
}

@inproceedings{akyildiz2025ambient,
	title={Deep Learning for Multi-Tag Ambient Backscatter Communications},
	author={Aky{\i}ld{\i}z, Talha and Mahdavifar, Hessam},
	booktitle={Proc. 59th Asilomar Conf. Signals, Syst., and Comput.},
	address = {Pacific Grove, CA, USA}, 
	month={Nov.},
	year={2025},
}
	
\end{document}